\documentclass[10pt,letterpaper]{article}
\usepackage[top=0.85in,left=2.75in,footskip=0.75in]{geometry}

\usepackage{amsmath,amssymb}

\usepackage{changepage}
\usepackage{filecontents}

\usepackage[utf8x]{inputenc}

\usepackage{graphicx}
\usepackage{caption}
\usepackage{subcaption}
\usepackage{rotating}
\usepackage{pdflscape} 

\usepackage{textcomp,marvosym}

\usepackage{cite}

\usepackage{nameref,hyperref}

\usepackage[right]{lineno}

\usepackage{microtype}
\DisableLigatures[f]{encoding = *, family = * }

\usepackage[table]{xcolor}

\usepackage{array}

\newcolumntype{+}{!{\vrule width 2pt}}

\newlength\savedwidth

\raggedright
\usepackage[aboveskip=1pt,labelfont=bf,labelsep=period,justification=raggedright,singlelinecheck=off]{caption}

\makeatletter
\renewcommand{\@biblabel}[1]{\quad#1.}
\makeatother

\date{}

\usepackage{lastpage,fancyhdr,graphicx}
\usepackage{epstopdf}
\fancyheadoffset[L]{2.25in}
\fancyfootoffset[L]{2.25in}
\begin{document}
\vspace*{0.2in}

\begin{flushleft}
{\Large
\textbf\newline{A method for assessing the success and failure of community-level interventions in the presence of network diffusion, social reinforcement, and related social effects} 
}
\newline
\\
Hsuan-Wei Lee\textsuperscript{1*},
G. Robin Gauthier\textsuperscript{1},
Jerreed D. Ivanich\textsuperscript{1},
Lisa Wexler\textsuperscript{2},
Bilal Khan\textsuperscript{1},
Kirk Dombrowski\textsuperscript{1}
\\
\bigskip
\textbf{1} Department of Sociology, University of Nebraska-Lincoln, USA
\\
\textbf{2} School of Public Health and Health Sciences, UMass Amherst, USA
\\
\bigskip

%
%





* hlee21@unl.edu

\end{flushleft}
\section*{Abstract}
Prevention and intervention work done within community settings often face unique analytic challenges for rigorous evaluations. Since community prevention work (often geographically isolated) cannot be controlled in the same way other prevention programs and these communities have an increased level of interpersonal interactions, rigorous evaluations are needed. Even when the `gold standard' randomized control trials are implemented within community intervention work, the threats to internal validity can be called into question given informal social spread of information in closed network settings. A new prevention evaluation method is presented here to disentangle the social influences assumed to influence prevention effects within communities. We formally introduce the method and it's utility for a suicide prevention program implemented in several Alaska Native villages. The results show promise to explore eight sociological measures of intervention effects in the face of social diffusion, social reinforcement, and direct treatment. Policy and research implication are discussed.



\section*{Introduction}
Assessments of the success and failure of interventions in a \textit{community} (rather than \textit{clinical}) setting tend to focus on individual behavior change or in the case of protective factors, behavior maintenance, normally by comparing the outcomes of intervention participants with those of non-participants \cite{spector_research_1981}. Social forces are recognized to influence those outcomes in two ways. In the first, participant/non-participant social context are thought to potentially influence the efficacy of the intervention---raising or lowering the impact on the basis of factors beyond the control of those performing it. To account for this possibility, analyses of outcomes generally depend on the use of ``control variables'' in statistical sense, i.e. measures such as gender or income or age that are assumed to serve as adequate stand-ins or predictors for the actual social position of the individual participant or non-participant and which may confound intervention results. Of course, the extent to which things like gender and age and income are in fact adequate stand-ins for that individual's social position is seldom tested, and it is not altogether clear how they might be considered when our purpose is to examine specific individual behaviors.  

The second way that social factors may influence the outcome of an intervention is when close social ties between participants and non-participants create the possibility of information transfer or behavioral influence between them \cite{snijders_introduction_2010}. One can easily imagine a situation, for example, where participants of smoking cessation intervention have close family members who are smokers. If the former seek to influence the behavior of the latter, then community wide outcome measures will be subject to ``contamination''. For this reason, most clinical trials require randomized control trials or RCT's. Here participants and non-participants are assumed to be isolated from one another, thereby preventing contamination from taking place. Such conditions are often unachievable in community-based intervention settings \cite{sanson-fisher_limitations_2007}. It is important to measure contamination, contagion, diffusion and a host of other transmission effects in community intervention settings where participants and nonparticipants are likely to be socially connected. Typically, these effects go un-noticed or un-measured thereby skew or confound results though to be more rigorous. Unfortunately, methods for measuring contamination and related social effects are scarce. 

In what follows, we propose a method for measuring the success and failure of behavioral interventions in small, relatively closed community settings. This method makes use of data on dyadic (network) ties to supplant more coarse control variables. Using dyadic data, we locate actors in their actual social position viz those around them, rather than using control variables as proxies of those positions. The method proposes four measures of the success or failure of the behavior in creating behavior change (from a state of absence to a state of presences) at a later time) and four measures of the success or failure of the intervention in promoting the maintenance of already existing protective behaviors (from a state of presence to a state of presences at a later time). It first tests both change and retention in an individual in the context of all social influences, and then uses similar strategies to assess the overall social impact of the intervention, the reinforcing effects of social relationships, and the overall diffusion of the intervention effect to non-participants.

To demonstrate the utility of the method, we employ a case study. The case study puts the full method in application on data from a recent suicide prevention intervention that includes mental health promotion, primary and secondary prevention components. Here we examine intervention efficacy across 39 protective behaviors, using post-intervention results from six communities in Alaska.

Prior attempts to deal with the possibility of contamination in a community setting can be found in the work of An, who proposes strategies such as novel forms of randomization \cite{an_multilevel_2015} and the strategies that take advantage of existing clustering \cite{an_peer_2011}. Similar strategies have been employed by Ugander \cite{ugander_graph_2013}, where by the authors propose a method to assess the effects of an A/B on-line experiment while being able to adjust for the effects of social influence.   Others have proposed straightforward means for measuring social influence, including Actor-Partner Independence models \cite{kenny_dyadic_2006} and various forms of binary outcomes assessment using multilevel logistic regression modeling \cite{mcmahon_guide_2006}. Others have proposed strategies for latent influence effects in a structural equation environment \cite{gonzalez_correlational_1999}. Here we propose a simple, flexible way to pose questions related to social factors that potentially arise in the context of intervention strategies in close-knit community contexts, and show how several of these social and relational factors can be analyzed for their effect on the intervention outcomes. The case study described below occurred in relation to a community intervention, PC CARES \cite{wexler2016creating,wexler2017promoting}.

\section*{Basic Definitions}
\label{sec:definitions}

To study the effect of a social intervention on a group of people with an underlying network structure, we compare ego-alter dyads at different time steps. Our study assumes that data collection proceeds at two time points, one before and one after intervention survey. On the underlying graph $G$ at two time points, we observe dyads of ego-alter pairs and examine the state transitions of dyads. 

Given a graph $G = (V, E)$, define the participate function $A(v): V \rightarrow \{0, 1 \}, \forall v \in V$, we have 
\begin{equation}
\label{A-def}
    A(v)=\left\{
                \begin{array}{ll}
                  0, \text{ if } v \text{ did not participate}\\
                  1, \text{ if } v \text{ participated.} 
                \end{array}
              \right.
\end{equation}
Also define the behavior function $B(v) \times  \{0, 1 \}: V \rightarrow \{0, 1 \}, \forall v \in V$ 
\begin{equation}
    B(v,i)=\left\{
                \begin{array}{ll}
                  0, \text{ if } v \text{ did not have the behavior at time } t_i\\
                  1, \text{ if } v \text{ had the behavior at time } t_i. 
                \end{array}
              \right.
\end{equation}
At two time points $t_0, t_1 \in \mathcal{R}, t_0 < t_1$, each node $v \in V$ is either in or not in the intervention in both times, but at time $t_0$ and $t_1$ separately, each node could either be with or without the behavior. For $u, v \in V$ and $e = (u,v) \in E$, let $C: \{0,1\} \times \{0,1\} \times \{0,1\} \times \{0,1\} \times \{0,1\} \times \{0,1\}  \longmapsto \mathcal{N}$ be a function that counts number of dyads defined as

\begin{equation}
\begin{split}
&C(x,y,p,q,r,s):=\\
&\sum_{e = (u,v)\in E} \big(A(u)-x+1\big) \cdot \big(A(v)-y+1\big)\cdot \\
&\big(B(u,0)- p+1\big) \cdot \big(B(v,0)-q+1\big) \cdot\\
&\big(B(u,1)-r+1\big) \cdot \big(B(v,1)-s+1 \big) 
\end{split}
\end{equation}

where 
$A$ is the participation function and $B$ is the behavior function. For example, $C(1,0,1,1,0,1)$ measures number of dyads satisfying: the ego is in the intervention (because $x = 1$); the alter is not in the intervention (because $y = 0$); the ego has the behavior at time $t_0$ (because $p = 1$); the alter has the behavior at time $t_0$ (because $q = 1$); the ego does not have the behavior at time $t_1$ (because $r = 0$); the alter has the behavior at time $t_1$ (because $s = 1$).

\begin{figure}[!htbp]
\centering
    \includegraphics[width=\textwidth, height =.35\textheight]{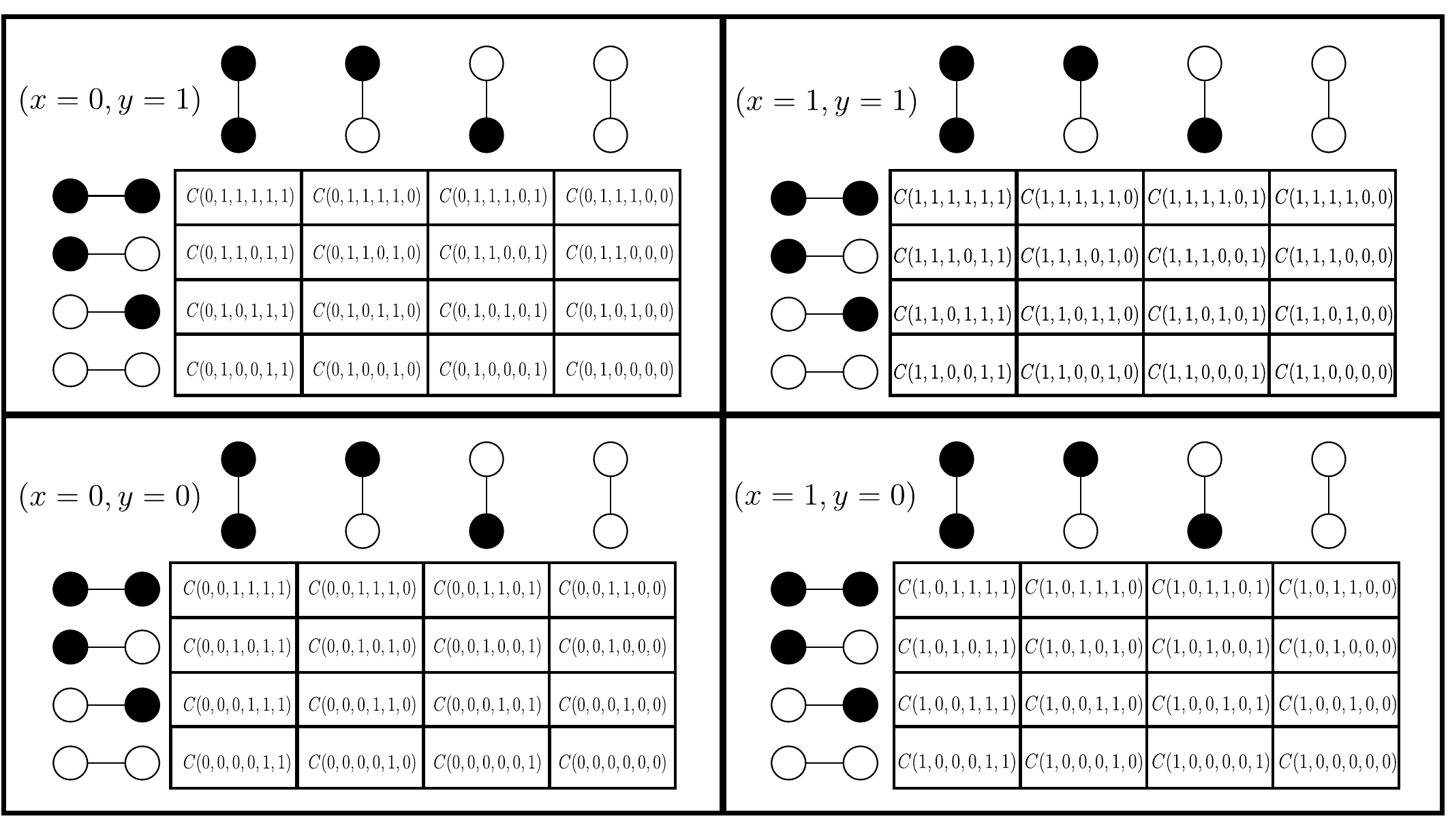} 
    \caption{Dyads behavior state transitions before and after the intervention. The four quadrants are the dyads of participation type $(x,y)$, the rows are dyads of $(p,q)$ at time $t_0$, and the columns are dyads of $(r,s)$ at time $t_1$. We color the nodes black if they have the behavior. For the horizontal dyads, the left nodes are ego and the right nodes are alters; for the vertical dyads, the top nodes are ego and the bottom nodes are alters.}
    \label{dyads64}
\end{figure}

With this counting notation, we use 6 bits to record the participation state and the behavior state at two time points of each end of the dyads. Each bit consists of a binary number, therefore we have $2^6 = 64$ types of dyads. The total 64 dyads are shown in Fig \ref{dyads64}.

Here we introduce more notations of counts. For each choice of $p,q \in \{0,1\}$ we define
\begin{equation}
_{0}N_{p,q} := \sum_{x,y,r,s \in \{ 0,1\}}C(x,y,p,q,r,s)
\end{equation}
which captures the number of dyads $(u,v)$ in which $u$ is of type $p$ and $v$ is of type $q$ at time $t_0$. For each choice of $r,s \in \{0,1\}$ we define
\begin{equation}
_{1}N_{r,s} := \sum_{x,y,p,q \in \{ 0,1\}}C(x,y,p,q,r,s)
\end{equation}
which captures the number of dyads $(u,v)$ in which $u$ is of type $r$ and $v$ is of type $s$ at time $t_1$. For each choice of $p,q,r,s \in \{0,1\}$ we define
\begin{equation}
N_{p,q,r,s} := \sum_{x,y \in \{0,1\}}C(x,y,p,q,r,s)
\end{equation}
which captures the number of dyads $(u,v)$ in which $u$ is of type $p$ and $v$ is of type $q$ at time $t_0$ and $u$ is of type $r$ and $v$ is of type $s$ at time $t_1$. Lastly, for each choice of $x,y,p,q \in \{0,1\}$ we define
\begin{equation}
_{0}N^{x,y}_{p,q} := \sum_{r,s \in \{0,1\}}C(x,y,p,q,r,s)
\end{equation}
which captures the number of dyads $(u,v)$ in which $(u,v)$ is of participation type $(x,y)$ and $u$ is of type $p$ and $v$ is of type $q$ at time $t_0$.

To understand the notations better, we consider some examples here. The notation $_{0}N_{1,0}$ captures the number of dyads $(u,v)$ in which $u$ has the behavior at time $t_0$ (because $p = 1$) and $v$ does not have the behavior at time $t_0$ (because $q = 0$). The notation $_{1}N^{1,0}_{0,1}$ captures the number of dyads $(u,v)$ in which $u$ is in the intervention (because $x = 1$), $v$ is not in the intervention (because $y = 0$), $u$ has the behavior at time $t_1$ (because $r = 1$), and $v$ does not have the behavior at time $t_1$ (because $s = 0$).

\section*{Eight Sociological Measures and Their Informal Interpretations}
\label{sec:measures}
Using this set of definitions, we formalize eight questions related to prevention (the maintaining of a protective behavior) and treatment (the adoption of a protective behavior previously absent) that are often difficult to answer when doing community intervention work---even when using ``gold standard" randomized control trials. The result of each test tell us whether we find a ``significant'' change in behavior, or in the case of protective behaviors, a ``significant'' continuation of already present protective behaviors.

\begin{enumerate}
\item \textbf{Direct Treatment Success in a Social Context} 
\\  Did the intervention promote a positive change in the behavior of someone who participated in PC-CARES \cite{wexler2017promoting} regardless of the intervention status or behavior change of their network ``alters''. We measure this by comparing the transitions from not having the behavior (or performing the activity) to having the behavior (or performing the activity) in those who were in the PC-CARES intervention vs. those who were not in the intervention, across all states of behavior/activity of their network alters.  

$$M1:=\sum_{y,q,s} C(1,y,1,q,0,s) - \sum_{y,q,s} C(0,y,1,q,0,s)$$ 

Our goal here is a basic measure of the success of the intervention on the participants given the full range of their social influences.
	
\item \textbf{Direct Prevention in a Social Context} 
\\Did the intervention promote sustained protective behaviors (or performance of a protective activity) in participants of the PC-CARES intervention regardless of the intervention status or behaviors/activities of their network alters. We measure this by comparing the maintenance of having the desired behavior in a PC-CARES participant versus  those who were not in the intervention, taking into account all of the social influences of their network alters. 

$$M2 := \sum_{y,q,s} C(1,y,0,q,0,s) - \sum_{y,q,s} C(0,y,0,q,0,s)$$ 

Our goal here is a measure of the success of the intervention in maintaining protective behaviors among participants given the broad mix of their social connections.
	
\item \textbf{Social Effect of Treatment} 
\\Did the participation of ego in the intervention induce a positive change in their alters' behavior, regardless of whether an alter participated in the intervention or not, AND whether the PC-CARES participant's behavior changed or not. We measure this by comparing the behavior change in the network alters of PC-CARES participants to the behavior change in the network alters of those who did not participate in PC-CARES. 

$$M3 := \sum_{y,p,r} C(1,y,p,1,r,0) - \sum_{y,p,r} C(0,y,p,1,r,0)$$ 

Our goal here is a measure of the general ``social effect'' of the intervention in promoting behavior change.
	
\item \textbf{Social Effect of Prevention}  
\\Did the participation of ego in the intervention induce a statistically significant continuation in their alters' protective behavior, regardless of whether that alter participated in the intervention or not, AND whether the PC-CARES participant's maintained the behavior or not. We measure this by comparing the continuation of protective behaviors in the network alters of PC-CARES participants to the continuation of protective behaviors in the network alters of those who did not participate in PC-CARES. 

$$ M4 := \sum_{y,p,r} C(1,y,p,0,r,0) - \sum_{y,p,r} C(0,y,p,0,r,0)$$

Our goal here is a measure of the general ``social effect'' of the intervention in promoting the maintenance of protective behaviors in the community as a whole. 
	
\item \textbf{Reinforcement of Change}  
\\Does it make a difference to the success of the treatment on PC-CARES participants when their network alters also participate in the PC-CARES intervention? We measure this by comparing behavior change of PC-Care participants whose network alters also participated with the behavior change of PC-CARES participants whose network alters did not participate. 

$$ M5 :=\sum_{q,s} C(1, 1, 1, q, 0, s) - \sum_{q,s} C(1, 0 , 1, q, 0 , s)$$

Our goal here is to measure whether there is a social reinforcement of the intervention effects created by having people participate in the intervention along with those with whom they have an already existing social relationship. 
	
\item \textbf{Reinforcement of Prevention} 
\\ Does it make a difference to the success of the  PC-CARES intervention in promoting the maintenance of protective behaviors among those who participate in the intervention when their network alters also participate in the PC-CARES intervention? We measure this by comparing the continuation of protective behaviors of PC-Care participants whose network alters also participated with the maintenance of protective behaviors of PC-CARES participants whose network alters did not participate. 
    
$$M6 := \sum_{q,s} C(1, 1, 0, q, 0, s) - \sum_{q,s} C(1, 0, 0, q, 0, s)$$

Our goal here is to measure whether there is a social reinforcement of the protective effects created by having people participate in the intervention along with those with whom they have an already existing social relationship. 
	
\item \textbf{Diffusion of Change}  
\\ How effective is the intervention for promoting behavior change among nonparticipants in PC-CARES when their network alters participated in PC-CARES but the subject did not. We measure this by comparing positive behavior changes in egos who did not participate but whose alters did participate to egos who did not participate and whose alters also did not participate. 

$$ M7 := \sum_{p,r} C(1, 0, p, 1, r, 0) - \sum_{p,r} C(0, 0, p, 1, r, 0)$$

Our goal here is measure the diffusion of the PC-CARES intervention treatment on non-participants with a direct link to a participant.
	
\item \textbf{Diffusion of Prevention}  
\\ How effective is the intervention on preserving protective behaviors among nonparticipants whose alters participated in PC-CARES? We measure this by comparing maintenance of protective behaviors in egos who did not participate but whose alters did participate with egos who did not participate and whose alters also did not participate. 

$$ M8 := \sum_{p,r} C(1, 0, p, 0, r, 0) - \sum_{p,r} C(0, 0 , p, 0, r, 0)$$

Our goal here is to measure the diffusion of the protective effects of the PC-CARES intervention on non-participants with a direct link to the a participant. 
\end{enumerate}

\begin{figure*}[!t]
  \begin{subfigure}[b]{0.5\linewidth}
  \centering
    \includegraphics[width=.75\linewidth]{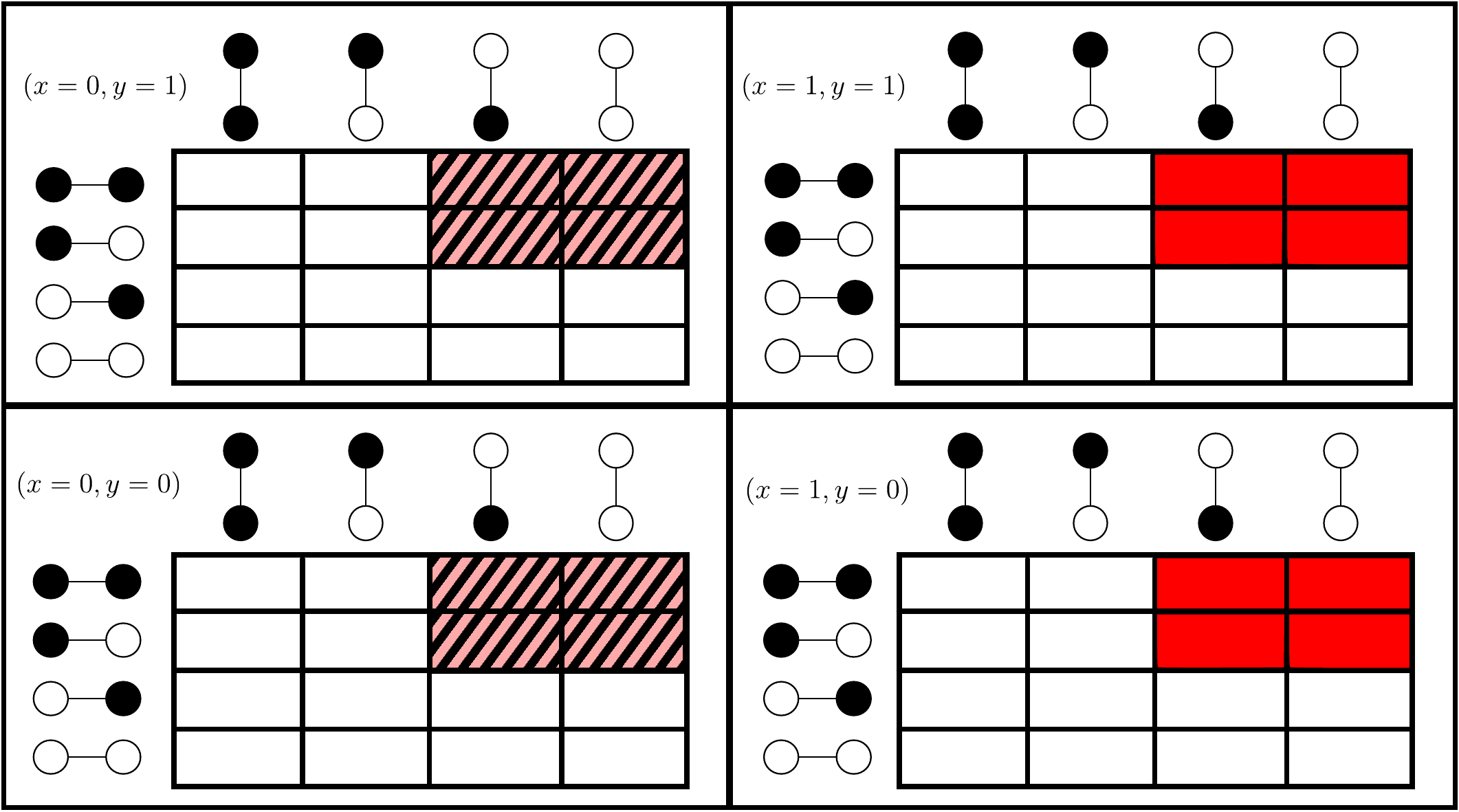}
    \caption{Measure $(M1)$}
  \end{subfigure} 
  \begin{subfigure}[b]{0.5\linewidth}
  \centering
    \includegraphics[width=.75\linewidth]{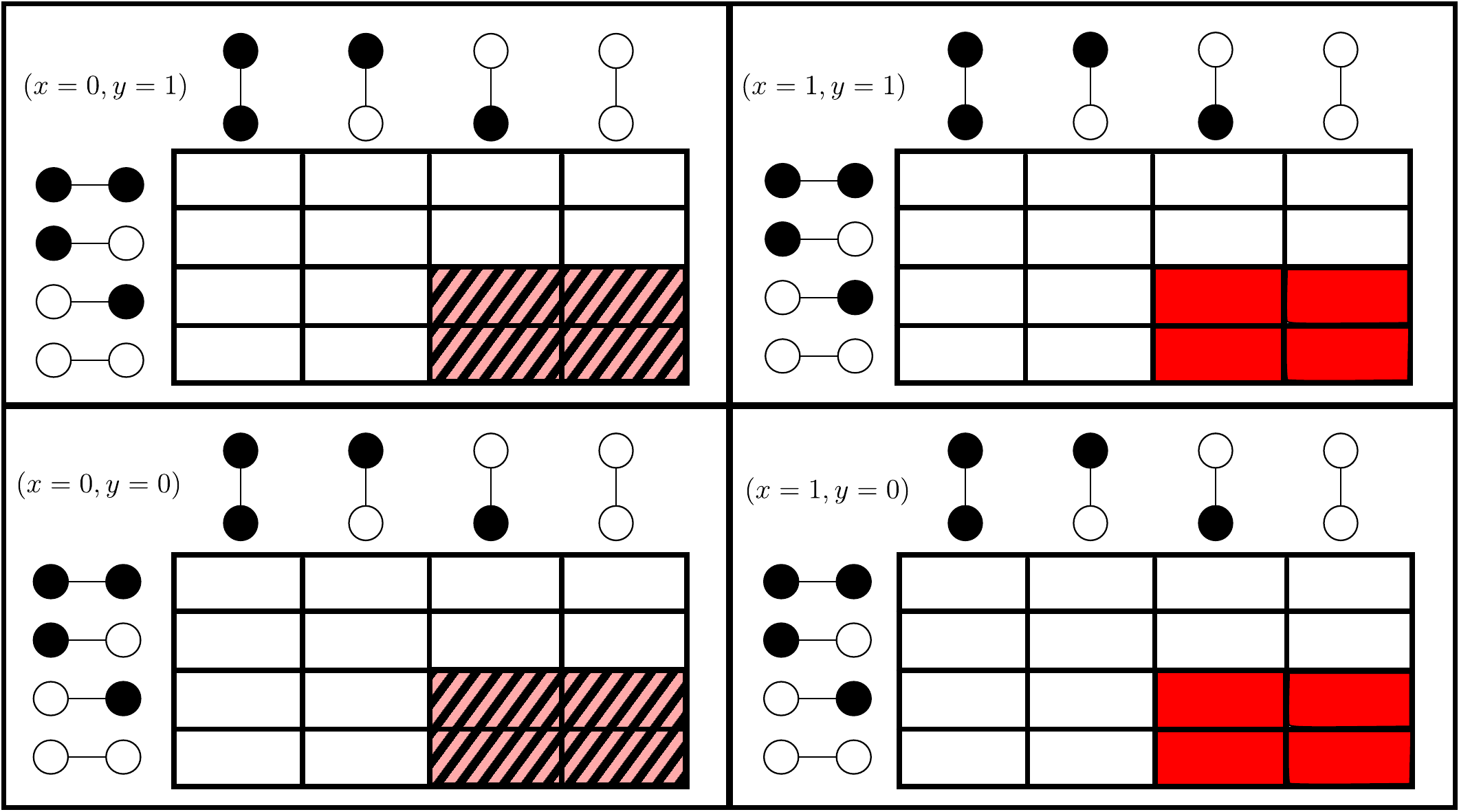}
    \caption{Measure $(M2)$}
  \end{subfigure}
  \begin{subfigure}[b]{0.5\linewidth}
  \centering
    \includegraphics[width=.75\linewidth]{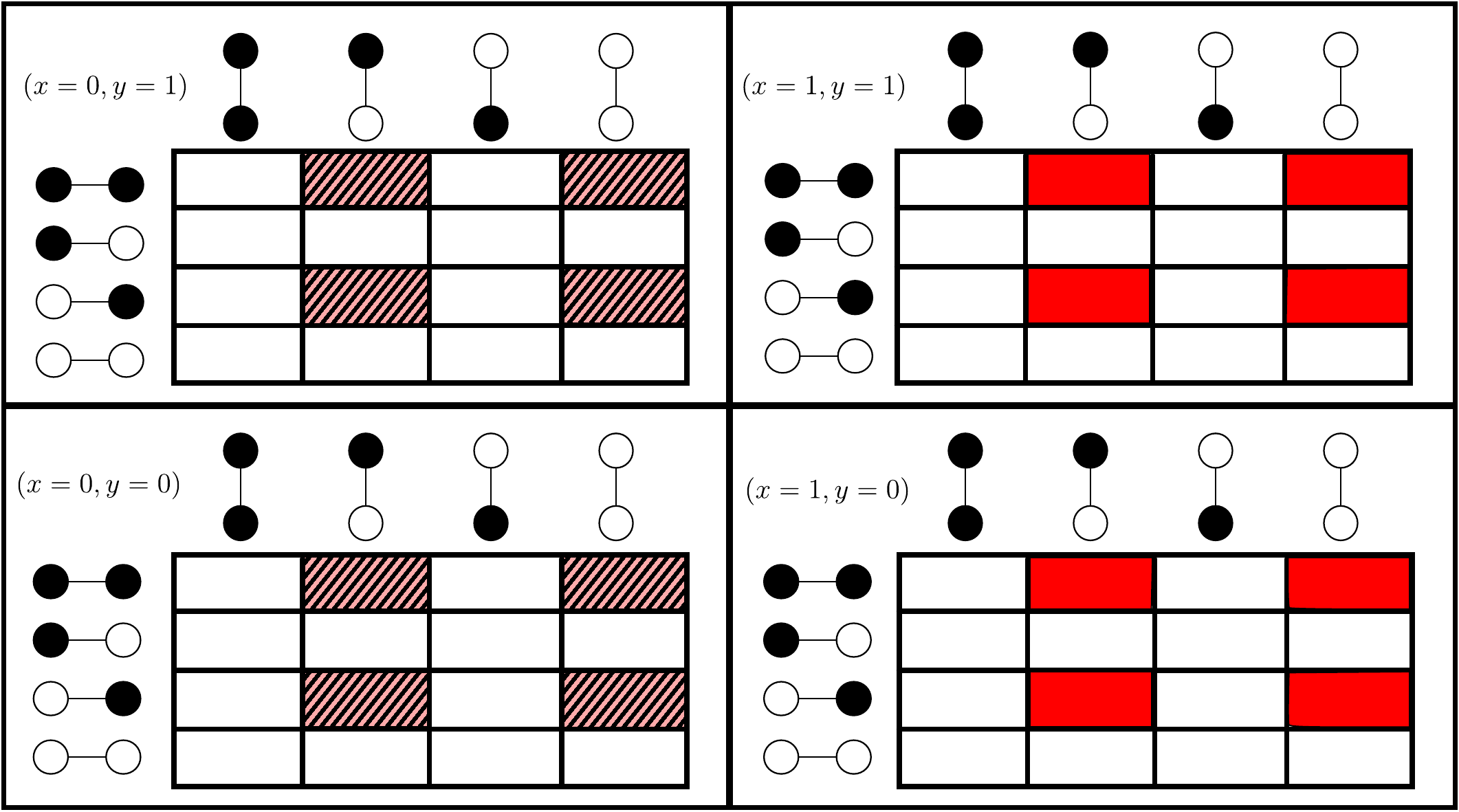}
    \caption{Measure $(M3)$}
  \end{subfigure}
  \begin{subfigure}[b]{0.5\linewidth}
  \centering
    \includegraphics[width=.75\linewidth]{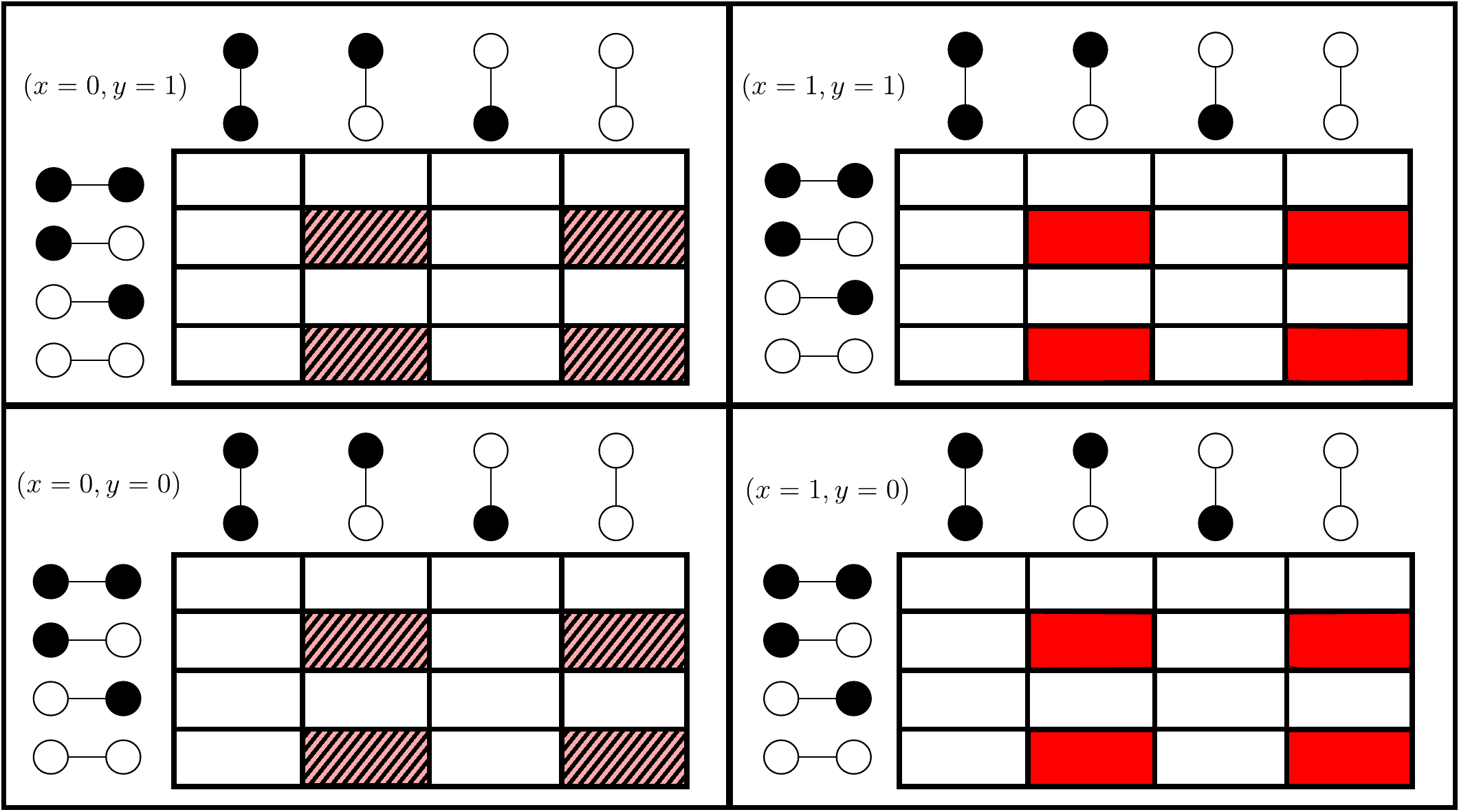}
    \caption{Measure $(M4)$}
  \end{subfigure}
  \begin{subfigure}[b]{0.5\linewidth}
  \centering
    \includegraphics[width=.75\linewidth]{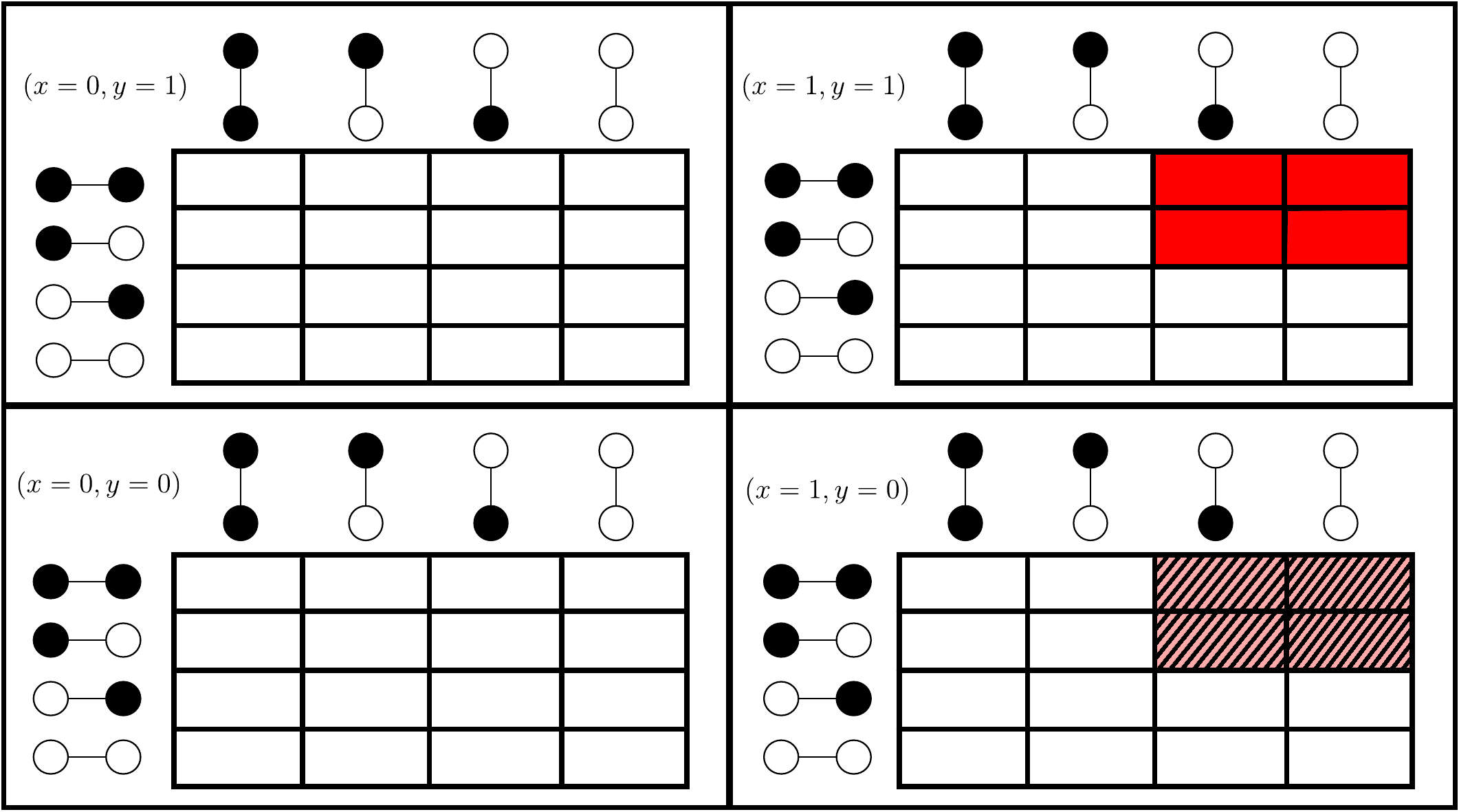}
    \caption{Measure $(M5)$}
  \end{subfigure}
  \begin{subfigure}[b]{0.5\linewidth}
  \centering
    \includegraphics[width=.75\linewidth]{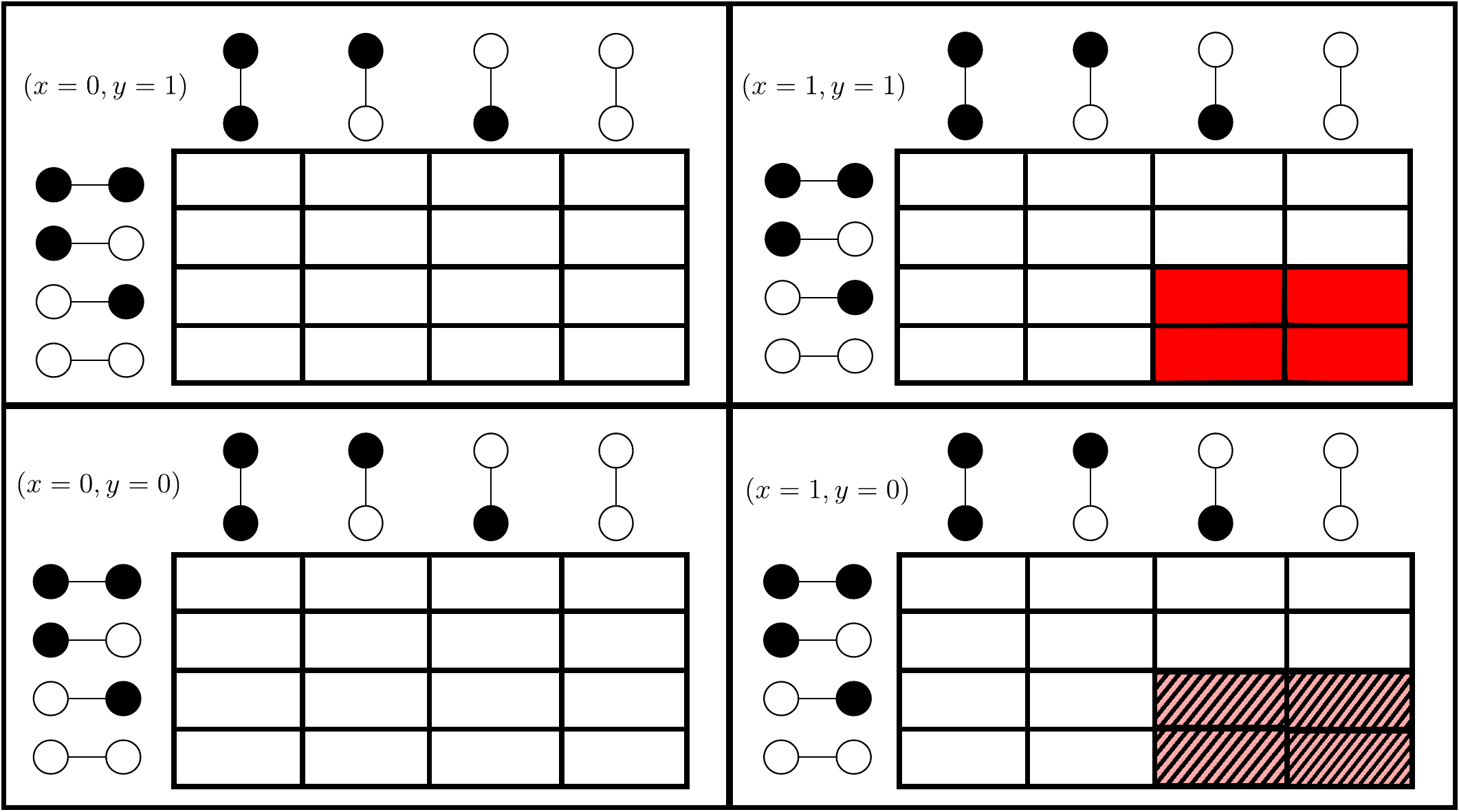}
    \caption{Measure $(M6)$}
  \end{subfigure}
  \begin{subfigure}[b]{0.5\linewidth}
  \centering
    \includegraphics[width=.75\linewidth]{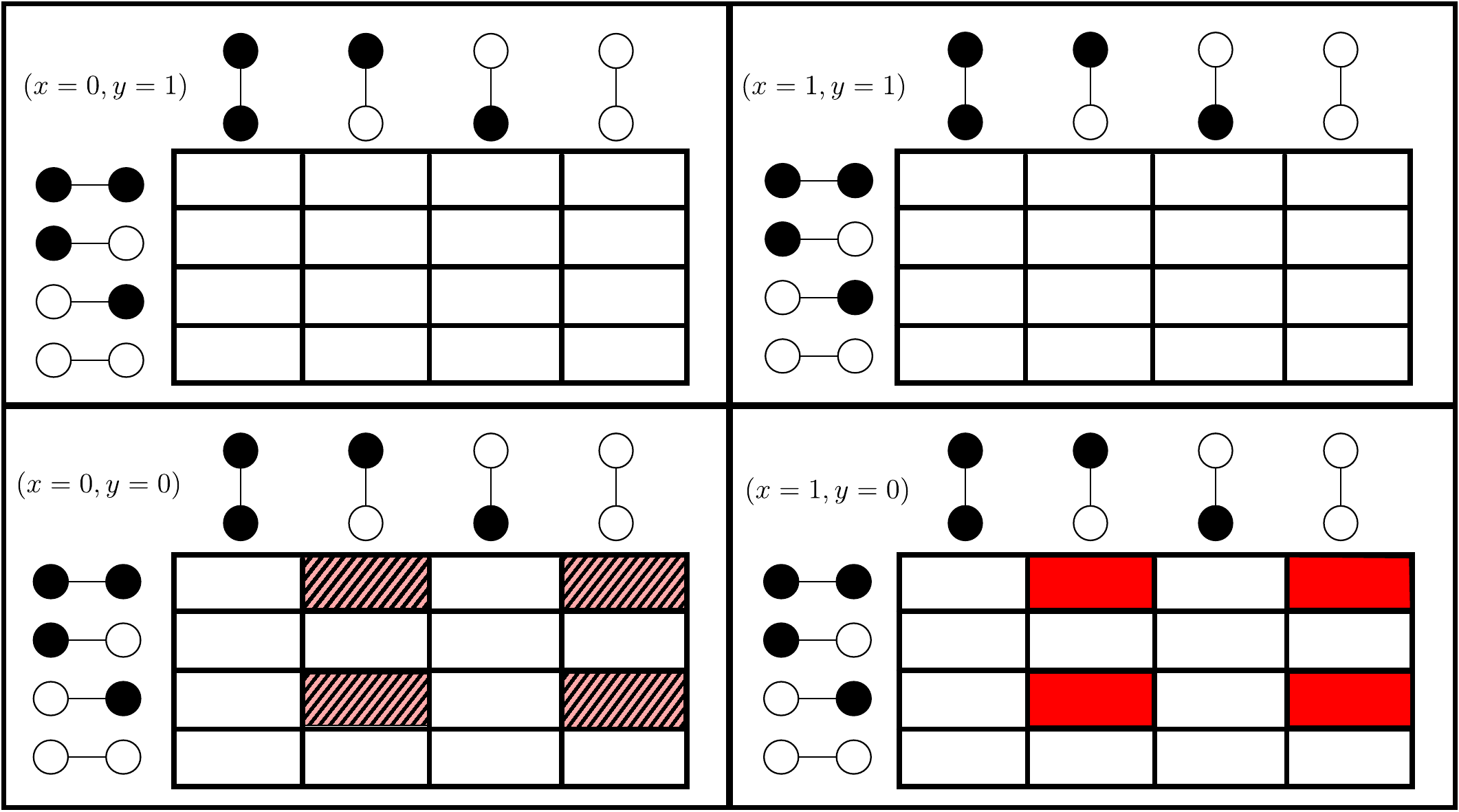}
    \caption{Measure $(M7)$}
  \end{subfigure}
  \begin{subfigure}[b]{0.5\linewidth}
  \centering
    \includegraphics[width=.75\linewidth]{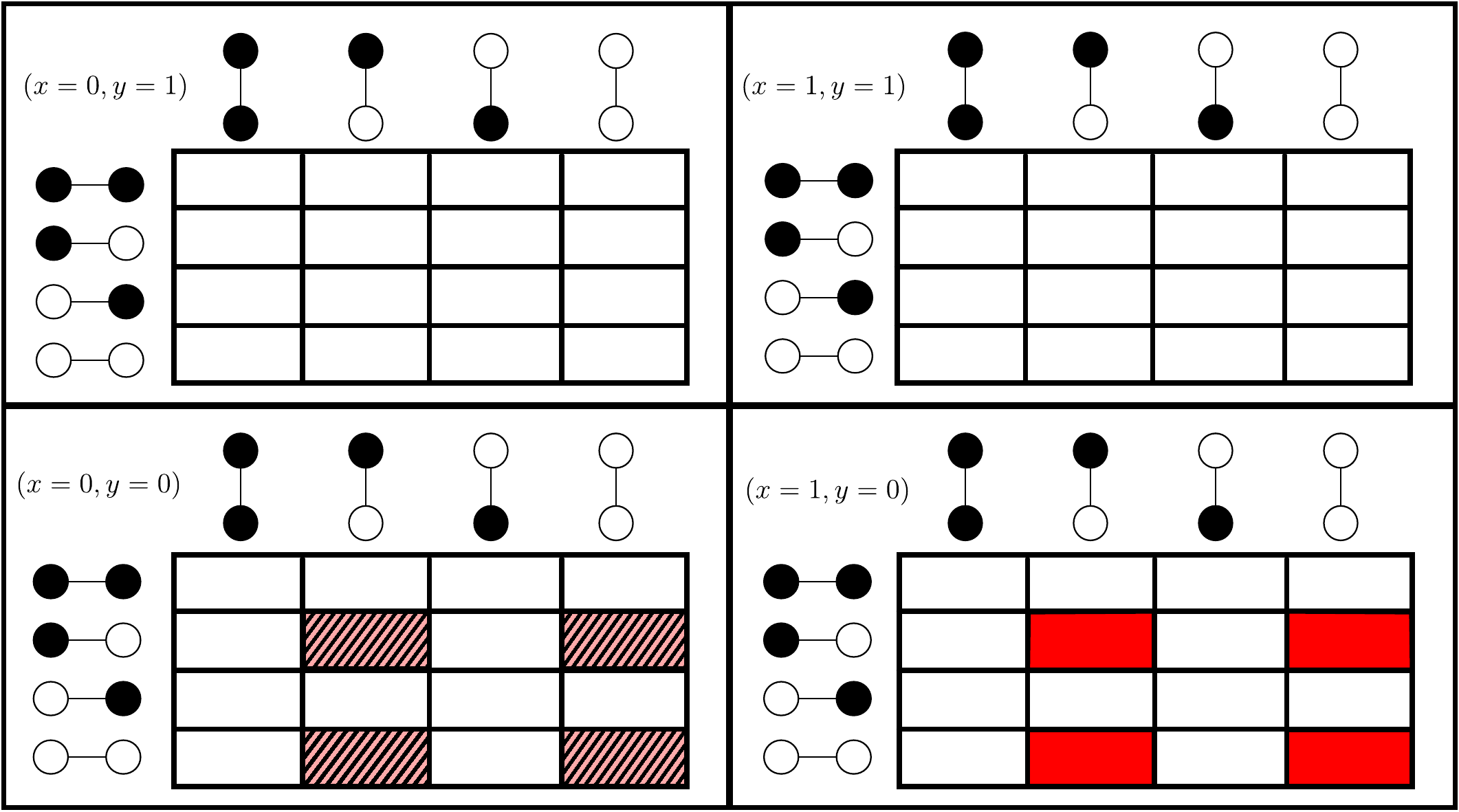}
    \caption{Measure $(M8)$}
  \end{subfigure}
  \caption{Geometry of measure decomposition. The four quadrants are the dyads of participation type $(x,y)$, the rows are dyads of $(p,q)$ at time $t_0$, and the columns are dyads of $(r,s)$ at time $t_1$. We color the nodes black if they have the behavior. For the horizontal dyads, the left nodes are ego and the right nodes are alters; for the vertical dyads, the top nodes are ego and the bottom nodes are alters. In each measure, the first sum is colored as red and the second sum is colored as pink with stripes.}
\label{geometry}
\end{figure*}

\section*{Investigating the effect of the intervention}

Our main goal is to quantify and compare the effect of social interventions of the ego and alter behavior by asking sociological meaningful measures. To consider all the possible measures of ego and alter related to the intervention and behavior, we separate the counting function $C(x,y,p,q,r,s)$ into two parts: participation and behavior parts. The participation part ($xy$ -part) cuts the probabilities into combinations of in and not in the intervention. The behavior part ($pqrs$ -part) cuts the probabilities into combinations of having or not having the behavior at different times $t_0$ and $t_1$.

To verify whether the dyads in measure display some feature with sociological meanings, we set up a null model that is used as a term of comparison. In the null model assumption, we assume that the intervention does not make any change of egos' and alters' behavior. Although there are possibly different numbers of people in each $(x,y)$ configuration, under the null model assumption, they share the same probability distribution of changing their behavior at time $t_1$. That is, we have 4 types of ego-alter participation combinations, e.g., $(x, y) = (0, 0), (0, 1), (1, 0), (1, 1)$, and they should all have the same probability transition matrix $P_{0}$ of the 16 $(p,q,r,s)$ types. Therefore our first step is to find out the probability transition matrix $P_{0}$.

For each $p,q,r,s \in \{0,1\},$ note that
$$N_{p,q,r,s} = \sum_{x,y \in \{0,1\}} C(x,y,p,q,r,s),$$
collapses the $(x,y)$ coordinates together under the null hypothesis. Moreover, define
\begin{equation}
P_{0} := P_{p,q,r,s} := \frac{N_{p,q,r,s}}{_{0}N_{p,q}}
\end{equation}
to be a $4 \times 4$ matrix which has columns $(p,q) = (0,0), (0,1), (1,0), (1,1)$ and has rows $(r,s) = (0,0), (0,1), (1,0), (1,1)$. Note that $P_{0}$ is a probability transition matrix and has row sum to be 1.  Still under the null hypothesis, we assume for each pair of $(x,y,p,q)$, it has a mass 
$$_{0}N^{x,y}_{p,q} = \sum_{r,s \in \{0,1\}} C(x,y,p,q,r,s).$$ This is the sum of people with pair of fixed $(x,y,p,q)$ values, and we use $_{1}N^{x,y}_{p,q}$ as the mass to generate the corresponding $C(x,y,p,q,r,s)$ under the null hypothesis based on the transition matrix $P_{0}$. Fix $x',y',p',q' \in \{0,1\}$, the null model yields 
\begin{equation}
I(x',y',p',q',r,s) = P_{p',q',r,s},
\end{equation}
where $r,s \in \{0,1\}$ and $I$ is the indicator function. Therefore, in the null model, we have $C(x',y',p',q',r,s)$ following a multinomial distribution with $_{0}N^{x',y'}_{p',q'}$ independent trials and the corresponding probabilities $P_{p',q',r,s}$, where $r,s \in \{0,1\}$. 

With the null model and the multinomial distributions computes from different sets of the observed
$C(x,y,p,q,r,s)$, we can use the bootstrapping technique to simulate the approximated distributions for each measure $(M1)-(M8)$ by sampling the numbers $B_{N}$ number of times. For a given data set with have a specific value of each measure $(M1)-(M8)$ as $d1$, ..., $d8$, we then compare each of them with the corresponding bootstrapped values $d1'$, ..., $d8'$ and see its tail probability of each measure. If the tail probability is less than some assigned significance level $\alpha$, then we have an $\alpha$-level of confidence to reject the null model assumption of the measure.

In addition to using the bootstrapping method to compute the tail probability of each measure, there is also analytical way of calculation. In each measure $(M1)-(M8)$, there are two summations of counts. Each count satisfies a multinomial distribution, hence the summations are sums of the multinomial distribution. The combined sum then satisfies a Poisson-Binomial distribution, which is a discrete probability distribution of a sum of independent Bernoulli trials that are not necessarily identically distributed. Suppose we have $N$ independent trials with probabilities f success and failure, for the $k$th trial, equal to $p_{k}$ and $1-p_{k}$, respectively. Let $X$ be the number of success in $N$ trials. The number of success $X = m$, or the probability mass function of $Pr(X = m)$ could be written as the sum
\begin{equation}
\label{PB-eq}
Pr(X = m) = \sum_{A \in F_{k}} \prod_{i \in A}p_{i} \prod_{j \in A^{c}} (1-p_{j}),
\end{equation}
where $F_{k}$ is the set of all subsets of $k$ integers that can be selected from $\{1,2,3,...,N\}$. Furthermore, the subtraction in each measure makes the final distribution satisfies the convolution of two Poisson-Binomial distributions by using the convolution formula
\begin{equation}
f_{X-Y}(z) = \sum^{\infty}_{x = -\infty} f_{X}(x) f_{Y}(x-z),
\end{equation}
where $X$ and $Y$ are two independent Poisson-Binomial distributions. Since we know the distribution of null model, the tail probabilities of the observed values could be computed accordingly. 

\section*{PC CARES Program Evaluation}

\subsection*{Background}
Suicide among Indigenous populations has remained a problem across American Indian populations living on the reservation \cite{walls_strain_2007}, and off the reservation \cite{ivanich_suicide_2017}. The problem, however, has remained a even more problematic for Alaska Native Youth \cite{allen_suicide_2009,wexler_factors_2012}. In line with the the recommendations of Okamoto \cite{okamoto_continuum_2014}, scholars working with Indigenous communities such as Allen et al. \cite{allen_protective_2014}, and Whitbeck \cite{whitbeck_guiding_2006} suggest that using prepackaged prevention programs are problematic when working with Indigenous populations, that stand in need of localized mobilization of culturally tailored programs adapted or created for targeted needs. 

To address concerns of suicide among ten Alaska Native villages, Promoting Community Conversations About Research to End Suicide (PC CARES) was created. The program is a community health education intervention created at the grassroots level with both general community feedback and targeted community leader (police, social work, local government officials, etc.) input. All curriculum and materials are created by the communities, for the communities, with the intent of leaving a sustainable and adaptable prevention program for years to come. Data for this study was collected among six of the ten villages. 

\subsection*{Data}
To illustrate the full method discussed in this paper, we use data from the Promoting Community Conversations About Research to End Suicide (PC CARES) project conducted in a remote and rural region of Alaska. PC CARES is a community-based suicide prevention program that mobilizes community members to use information from research to determine how they will reduce risk and increase protective factors for suicide among Alaska Native communities. Detailed discussion of the model, community engagement, and justifications for PC CARES is found here \cite{wexler2017promoting}. Collectively, PC CARES project was held in ten Alaska Native villages that share a similar culture and language. Of these participating communities, eight with the largest number of PC CARES participants were selected to participate in social network surveys. We collected data in eight villages. Five of the ten villages used respondent driven sampling (RDS) sampling strategy while the remaining three villages were small enough to achieve sufficient sampling with traditional random recruitment. In the  five villages that used a RDS sample design, there were 423 responses. In total, 447 individuals across the six selected communities participated in the data collection. 
 
With approval from the regional tribal health organization and the University of Massachusetts IRB, data collection for this novel method was done in six intervention villages by one or two researchers and their community partners. Before arriving in each village, announcements about the research let village members know that they could participate in a survey to identify what people are doing to help promote wellness in their lives and village. Importantly, research teams emailed and called those people who had participated in the intervention in order to maximize the number of PC CARES participants who filled out a survey. PC CARES participants were the first group recruited.

Before filling out the survey on iPad computers, participants went through a written consent process. They were told about the intervention research study, expectations of participation in the research, and possible risks and benefits. Focused on helping behaviors and support giving, the survey was minimal risk, and took about 20 minutes to complete. All participants received \$20 for their time, whether or not they competed the survey.

After each person completed the survey, s/he was given either two or three “coupons” to share with people close to them. The number of coupons given was related to the size of the village, and the level of participation in the data collection. Each “coupon” was worth 5 dollars if the person they gave it to came in to complete their own survey. Data collection continued for 2-3 days in each of the five communities where RDS was used. 

Individuals interviewed were asked a series of demographic characteristics, a series of 39 PC CARES program specific questions (see table \ref{tab:quest}), and were given a list of PC CARES participants and asked to identify if they were close to those on the list. In combination with the RDS edge list and the reported relationships in the individual surveys, a dyadic data set was created. Each dyad contained information on each person’s PC CARES completion status, and their before and after state on the program specific questions listed in table \ref{tab:quest}. The survey asked about respondent conversations, understandings and behaviors related to suicide prevention. All questions were binary yes or no if they exhibit the behavior in question at the before state and again in the after state.

\section*{Results}
\begin{figure} [!t]
\begin{adjustwidth}{-2.75in}{0in}
\centering
	\includegraphics[width=1.5\textwidth, height = .5\textheight]{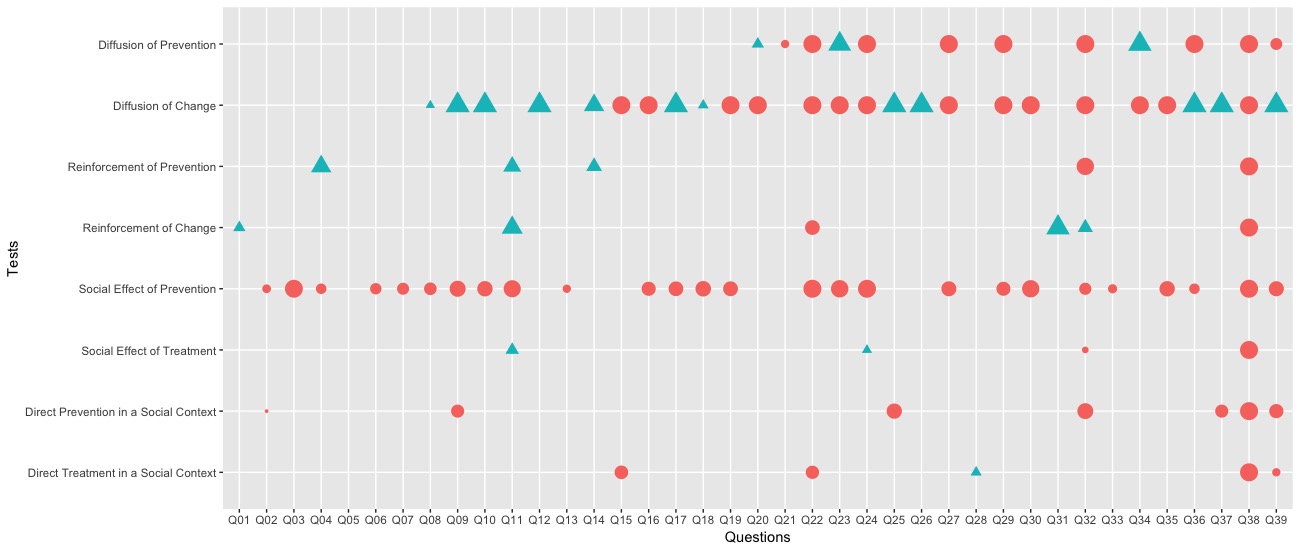}
    \caption{Significance Size and Directionality of Eight sociological Tests from PC CARES prevention program}
	\label{fig:bubble}
    \end{adjustwidth}
\end{figure}

Fig \ref{fig:bubble} displays the overall evaluation results across 39 items measured for pre and post intervention using the method proposed in this paper. Along the y-axis, eight sociological tests of program evaluation in a social context. Along the x-axis each of the 39 items measured are located. Circles suggest a positive change, where a triangle suggests that negative change was introduced as a result of the PC CARES program. The size of each symbol corresponds to how significant---or how far from the null model the real data was. The larger the symbol, the farther away the more significant the relationship. Intersections absent of circles or triangles suggests that no significant relations found for the item and sociological test. 

Overall, PC CARES seems to have introduced an overwhelming positive impact to the communities as noted with the number of circles. Specifically, 66/91 (73\%)  significant relationship found were positive changes in the community. However, 25 items across the eight sociological measures were found to have a significant negative relationship. Of the 25 items that were found to be a negative relationship, nearly half (12/25) were found for one specific sociological evaluation test ``diffusion of change''. Furthermore, the inclusion of item 11, ``Have you reached out to someone who attempted suicide?", accounts for 60\% (15/25) of all negative results are accounted for.    

In terms of sociological tests, PC CARES made the most impact in terms of \emph{social effect of prevention}. Across all 39 items, 26 (66.7\%) items showed significant positive change with zero negative impacts. Individuals that had an alter participate in the program was associated with a statistically significant continuation of protective factors regardless of their own participation in PC CARES. Simply put, it appears that the PC CARES program induced positive change without hands on participation. 

PC CARES had a large impact on two items of note. First, item 22, ``Have you had conversations about making it harder for an 'at risk' person to get a loaded gun?'' Four of the eight tests were statistically significant for creating positive change on this item. Secondly, item 38, ``Have you done something (subsistence, basketball, other activities) with a youth in the summer?'' Item 38 showed statistical significance for positive change on all eight sociological tests. 

Lastly, item 32, ``Have you taken action, like removing guns or alcohol, to make a home safer?'' also appears to have also been positively impacted by the PC CARES program. Although, this item showed significant negative change for reinforcement of change, it showed positive change on six other sociological tests. 

In sum, PC CARES, using this evaluation method, has made a overwhelming positive change to the communities it served. Items related to gun safety also seemed to benefit from the PC CARES program. Rigorous results of this form are often difficult to disentangle in isolated communities that rely on kinship and social interaction that are not often seen as being problematic in Eurocentric communities.

\section*{Conclusions}

Stanley and colleges \cite{stanley2017imperative} note the historical and contemporary difficulties establishing scientifically rigorous assessments in community prevention work. Opposed to clinical research, doing prevention evaluation within communities (often geographically isolated and heavily dependent on kinship) sits outside a controlled environment. The lack of exact control, presences of social dynamic process, and inclusion of cultural sharing systems establishes and environment that is often prone to critics against empirical assessments of community prevention work. 

This paper presents a method for assessing, not only the success, but the failures of community-level intervention in the presence of complex social relationships. As illustrated in this article, the proposed method, using of dyadic data, can be used to explore eight sociological measures of prevention outcomes in community-intervention work that are often difficult to assess even when ``gold standard'' randomized control trials are employed. The method presented here allows for exploration of prevention effects in the wake of social diffusion and social reinforcement of treatment. 

Investigators well situated to benefit from such a method include investigators that do prevention work among specialized populations, geographically isolated communities, and target populations where social interactions are assumed to occur naturally (and cannot be controlled). For instance, as the data suggests in this paper, investigators working with Alaska Native, American Indian, or Native Hawaiian can greatly benefit from this method---although these are only several examples. The benefits of using this method can equally extend to non-indigenous prevention scholarship, such as the current efforts among Appalachian populations \cite{hambymeaning} or school wide intervention efforts \cite{valente_peer_2007}.

\subsection*{Limitation and Future Research}

In spite of the contributions this paper provides to investigators doing prevention work within community settings, there are few key limitations that should be noted. First, the use of the proposed method requires the collection of social relations. While collecting relational data is not an inherit limitation, an added burden is placed on the researchers and respondent to know the prevention program status and behavior state of the respondents nominations. Second, to implement this method investigators should be sensitive to the thresholds required to ``fill" the cells shown in Fig \ref{dyads64}. For the method to be used, it is essential that counts of dyads are present for each cell. Lack of sufficient cell size may eliminate or reduce the sociological measures that can be assessed. 

It is also important to note that the data presented here may present some sampling limitations. The data used for this study come from geographically isolated communities located in rural Alaska. Potential confounds and social complexities may be present in studies that work with populations that can more-freely access other study sites. Additionally, when doing prevention work, one should also be cognizant of recruitment and participation self-selection bias. These bias may include treatment effects, but also the bias present in social nominations collected. 

Lastly, our analyses indicate that question 38, ``Have you done something (subsistence, basketball, other activities) with a youth in the summer?'' was significantly improved across all sociological measures. Here, we caution the interpretation of this result as this finding may be a product of measurement error. Given the timing of data collection for these communities, it is likely that engaging in outdoor activities and/or culturally specific activities, in optimal weather may have established a socially desirable response. Thus, this method is not robust against measurement error. 

Notwithstanding these limitations, the current study makes make several contributions to scholarship on program assessment and prevention science. Given the findings presented here, future research would benefit by building on the work presented here in several ways. First, the sociological measures presented here are not an exhaustive list of potential social and dynamic processes that investigators may care to assess and more work is needed to expand the list of social processes of interest among prevention science assessment. Second, this method should be used across multiple modes of study design to assess the robustness of the method when, unlike here, an RDS study design is not used. Finally, scholars should employ this method among a variety of populations, social context, and geographic locations to evaluate the effectiveness of such a method in cases that are not as kinship based---unlike the data presented here. 




\section*{Acknowledgments}
Cras egestas velit mauris, eu mollis turpis pellentesque sit amet. Interdum et malesuada fames ac ante ipsum primis in faucibus. Nam id pretium nisi. Sed ac quam id nisi malesuada congue. Sed interdum aliquet augue, at pellentesque quam rhoncus vitae.

\nolinenumbers


\clearpage

\section*{Appendix 1 - PC CARES Intervention Questions}
\begin{table}[!htbp]
\begin{adjustwidth}{-2.25in}{0in}
	\centering
		\caption{PC CARES Intervention Specific Questions}
	\begin{tabular}{cl}
		\hline    \textbf{\#} & \multicolumn{1}{c}{\textbf{Question}} \\
		\hline
		1     & Have you talked about the impact of culture loss on the lives of young people in your community? \\
		2     & Have you talked about how youth suicide attempts happen more often in the summer? \\
		3     & Have you talked with others about how showing you care can reduce the risk of suicide? \\
		4     & Have you talked with others about how to prevent suicide? \\
		5     & Have you talked with others about history and suicide? \\
		6     & Have you talked about how to give support for a person who was feeling low or suicidal? \\
		7     & Have you talked with someone about how culture can promote youth wellness? \\
		8     & Have you talked to others in your family about wellness? \\
		9     & Have you made efforts to talk more to a young person that you know? \\
		10    & Have you gotten ideas about how to support people close to you? \\
		11    & Have you reached out to someone who attempted suicide? \\
		12    & Have you recognized how to make positive change within your own family? \\
		13    & Have you done something to make someone feel cared about after a suicide attempt? \\
		14    & Have you helped in some way when noticing someone is having a hard time? \\
		15    & Have you done something to make a young person feel cared about? \\
		16    & Have you opened up to hear others? \\
		17    & Have you tried to listen more to a young person that you know? \\
		18    & Have you spent time listening to someone who just wanted to talk about their experience? \\
        19	  & Have you showed you cared just by hearing what someone had to say? \\
		20    & Have you showed you cared just by hearing what someone had to say? \\
		21    & Have you spoken up on what community organizations can do to reduce risk of youth suicide? \\
		22    & Have you had conversations about making it harder for an 'at risk' person to get a loaded gun? \\
		23    & Have you opened up to share your thoughts? \\
		24    & Have you spoken up about community protective factors? \\
		25    & Have you trusted others in the community to hear what you have to say? \\
		26    & Have you increased safety, like removing guns or staying with a person, when worried they might be suicidal?\\
		27    & Have you done something for prevention when worried about someone's risk of suicide? \\
		28    & Have you NOT talked about the details of a suicide for fear of increasing suicide? \\
		29    & Have you only talked about a suicide in a safe way? \\
		30    & Have you talked about suicide prevention? \\
		31    & Have you talked about how honoring a person who died by suicide? \\
		32    & Have you taken action, like removing guns or alcohol, to make a home safer?  \\
		33    & Have you let people know what resources are available for prevention? \\
		34    & Have you participated in wellness activities? \\
		35    & Have you taken a young person to do subsistence activities during the summer? \\
		36    & Have you suggested ways community organizations could work together to increase wellness?  \\
		37    & Have you talked with community members about wellness? \\
		38    & Have you done something (subsistence, basketball, other activities) with a youth in the summer? \\
		39    & Have you worked with others to increase wellness in the village? \\
		\hline
	\end{tabular}%
	\label{tab:quest}%
    \end{adjustwidth}
\end{table}%

\clearpage

\section*{Appendix 2 - Validation of the Bootstrapping Method and Direct Computation of the Poisson-Binomial Distributions} 

\begin{figure}[!ht]
 \centering
    \includegraphics[width=.5\textwidth]{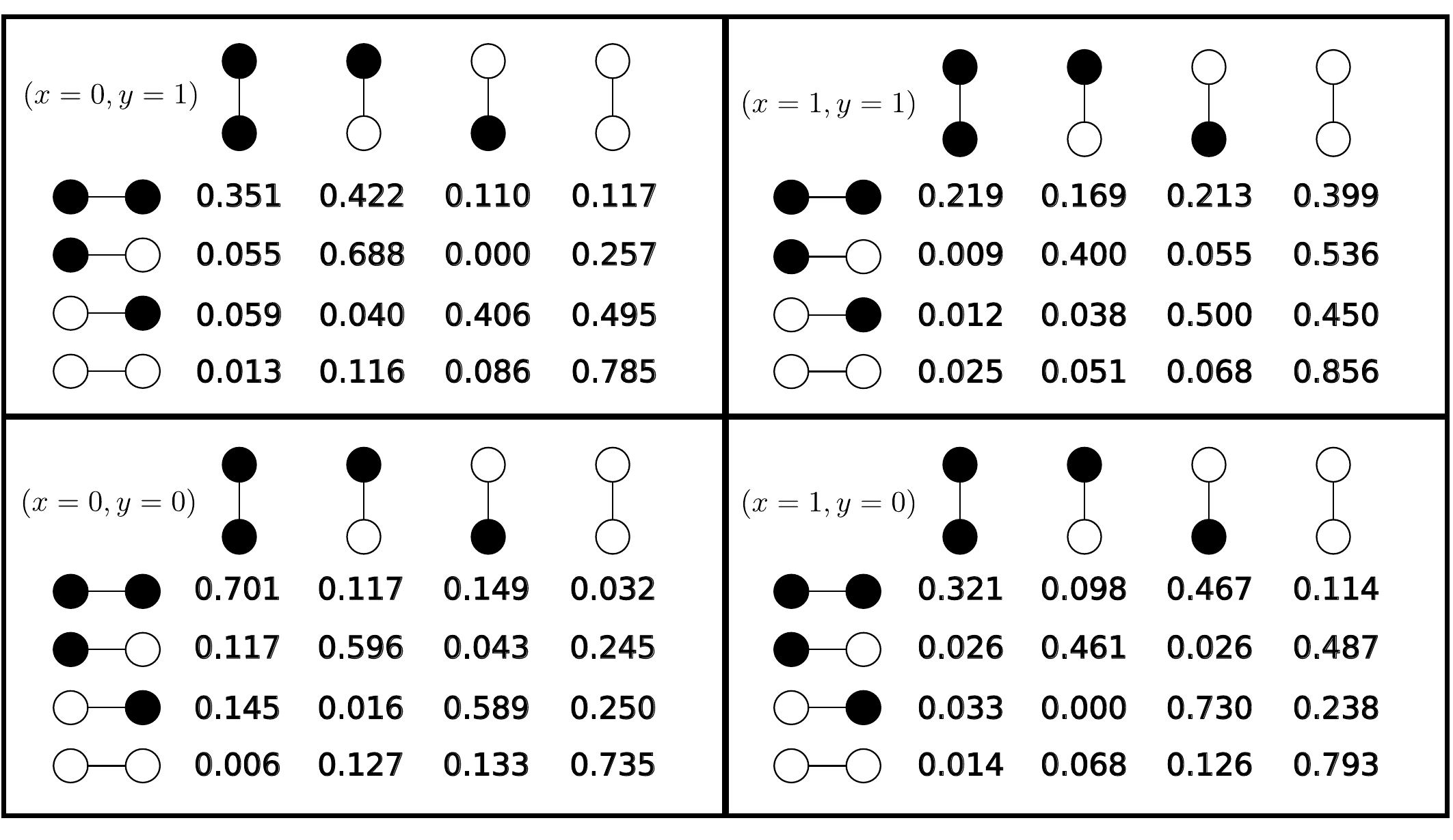}
    \caption{Simulated data for a plausible representation of a community based intervention. The four quadrants are the dyads of participation type $(x,y)$, the rows are dyads of $(p,q)$ at time $t_0$, and the columns are dyads of $(r,s)$ at time $t_1$. We color the nodes black if they have the behavior. For the horizontal dyads, the left nodes are ego and the right nodes are alters; for the vertical dyads, the top nodes are ego and the bottom nodes are alters.}
    \label{robin_data}
\end{figure}

The data presented in Fig \ref{robin_data} was created through a simulation designed to produce a plausible representation of a community based intervention designed to reduce the prevalence of a targeted behavior. We based the structure of social relations in our simulation on the friendship network of 444 middle school children. Using this network structure, each actor in the network was assigned an intervention participation status ($P(\{A(v) = 1\}) = 0.5, \forall v \in V$, see Eq \ref{A-def}) and an initial 0/1 behavior state was assigned stochastically based on the behavior states of their contacts (see Friedkin and Johnsen, 2011). The assignment of post-intervention states took place in two stages. An initial post-intervention behavioral state was generated for each node based on intervention participation. Actors who were randomly assigned to participate in the intervention had a low probability of a transition into the suppressed behavior (0→1: $p_{against}=0.05$) and a moderately high probability of transitioning out of that behavior state (1→0: $p_{effect}=0.5$). Actors assigned to the non-participation (control) group were assigned an equally low probability of transition into a present behavior state (0→1: $p_{stay}=0.05$), but given a much lower probability of transitioning into an absent one (1→0: $p_{change}=0.05$). Next, we update the post-intervention behavior state. Low to moderate levels of social conformity were enforced by allowing actors a twenty percent chance ($p_{social}=0.2$) to reassign their initial post-intervention behavior state to the state that is most common among their alters, based on these alters’ initial post-intervention value.

The simulation procedure produced 2397 dyads. Within each simulated dyad, egos and alters each possess a behavioral attribute characterized by two states; present or absent before and after the intervention, yielding four distinct dyadic states: (1) both ego and alter report a present behavior state; (2) only ego reports a present behavior state, while alter's state is absent; (3) ego’s state is absent and alter's is present; or (4) ego and alter report an absent state. Dyads are further divided into four groups based on the intervention status of both the ego and the alter. Within each group, a state transition matrix records counts of ego and alter's joint (dyadic) post-intervention state according to its pre-intervention state. The four combinations of pre-intervention dyadic behaviors are graphically depicted along the rows of each scenario, while post-intervention states are depicted along the columns. The probability of movement from pre to post-intervention states is conditioned on the pre-intervention status of the dyad, thus all rows sum to 1. 

\begin{figure*}[!tbp]
  \begin{subfigure}[b]{0.5\linewidth}
    \includegraphics[width=.75\linewidth]{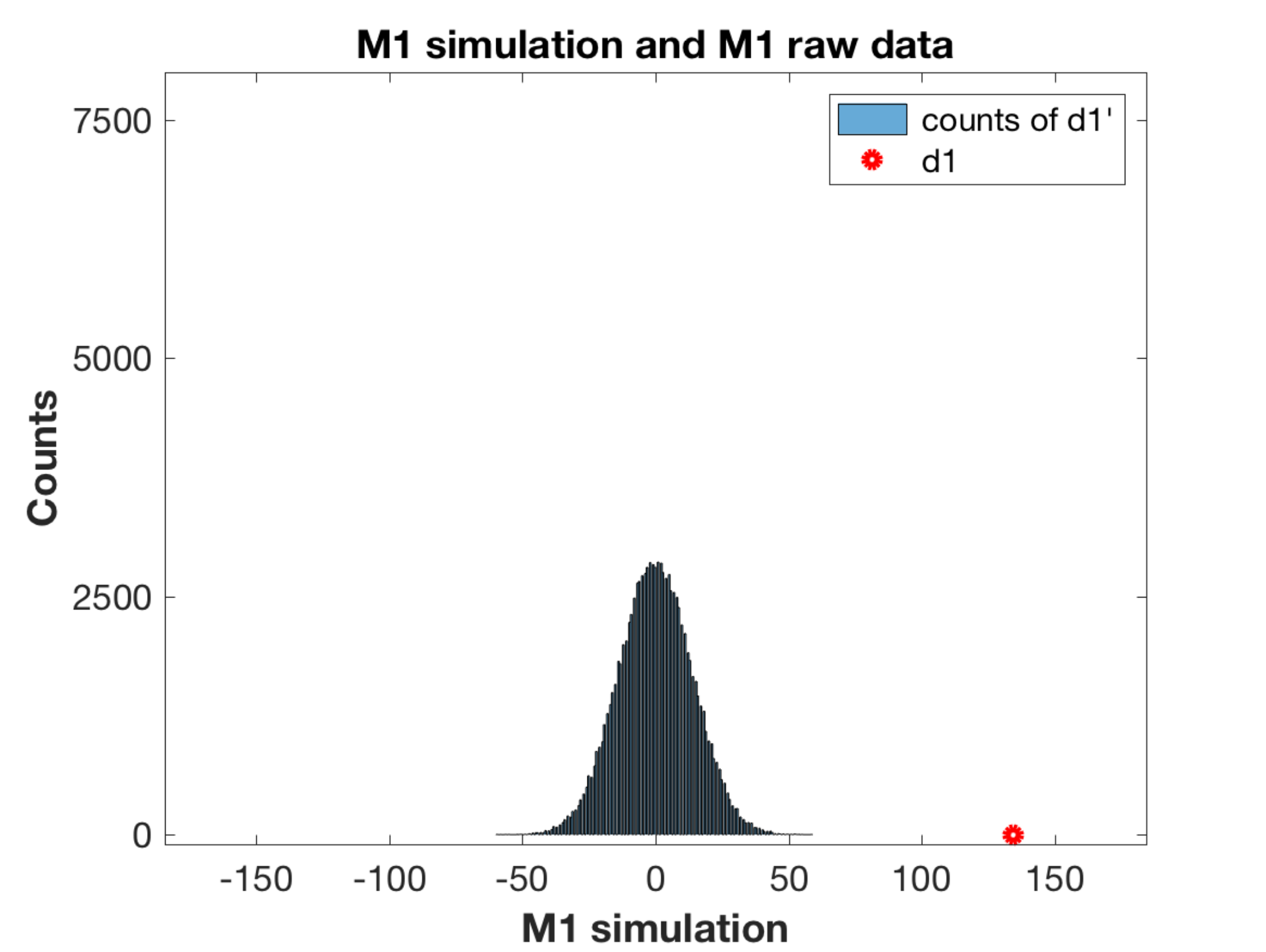}
    \caption{Measure $(M1)$, $\text{p-value} = 0$}
  \end{subfigure} 
  \begin{subfigure}[b]{0.5\linewidth}
    \includegraphics[width=.75\linewidth]{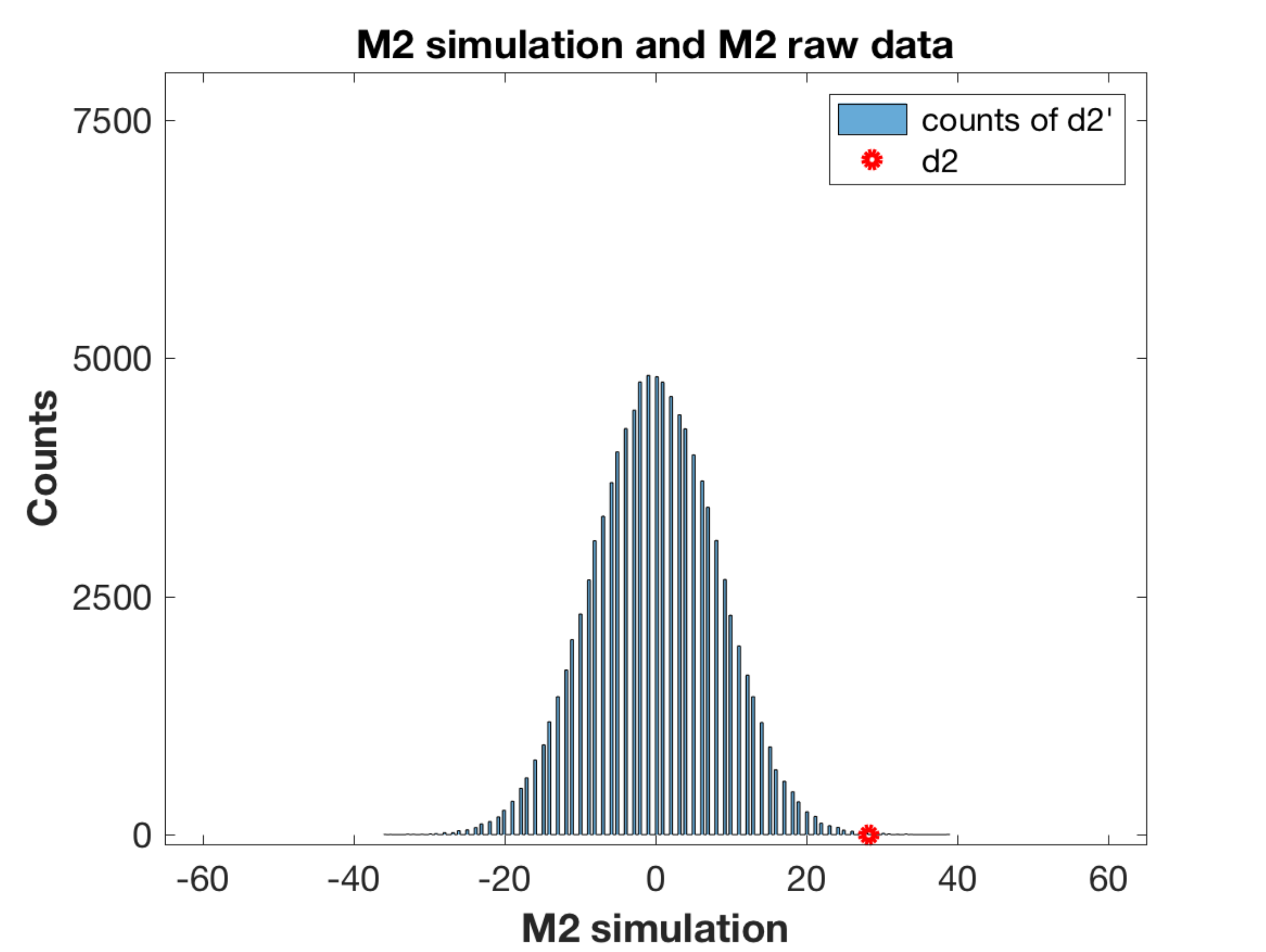}
    \caption{Measure $(M2)$,  $\text{p-value} = 0.003$}
  \end{subfigure}

  \begin{subfigure}[b]{0.5\linewidth}
    \includegraphics[width=.75\linewidth]{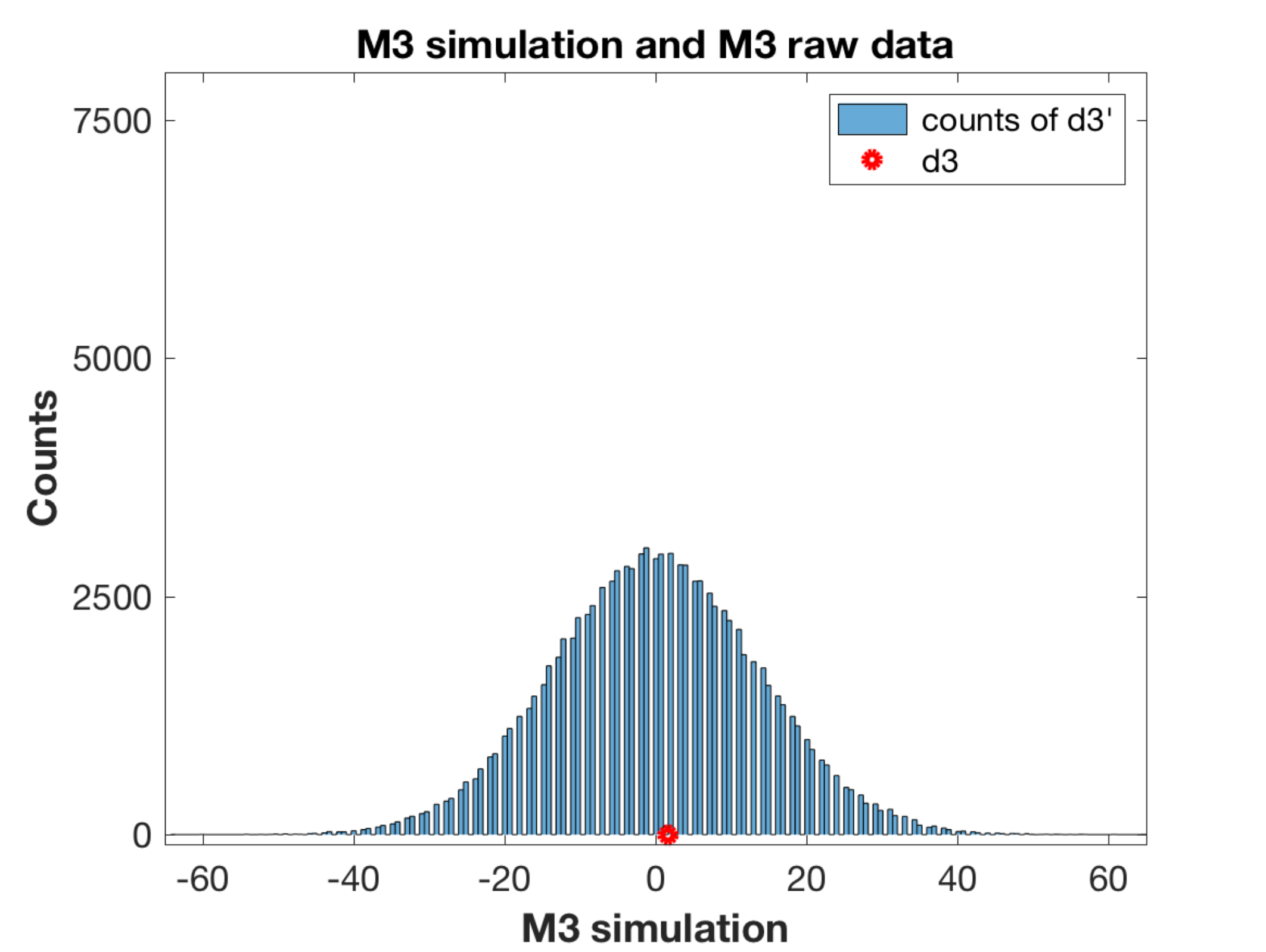}
    \caption{Measure $(M3)$, $\text{p-value} = 0.4560$}
  \end{subfigure}
  \begin{subfigure}[b]{0.5\linewidth}
    \includegraphics[width=.75\linewidth]{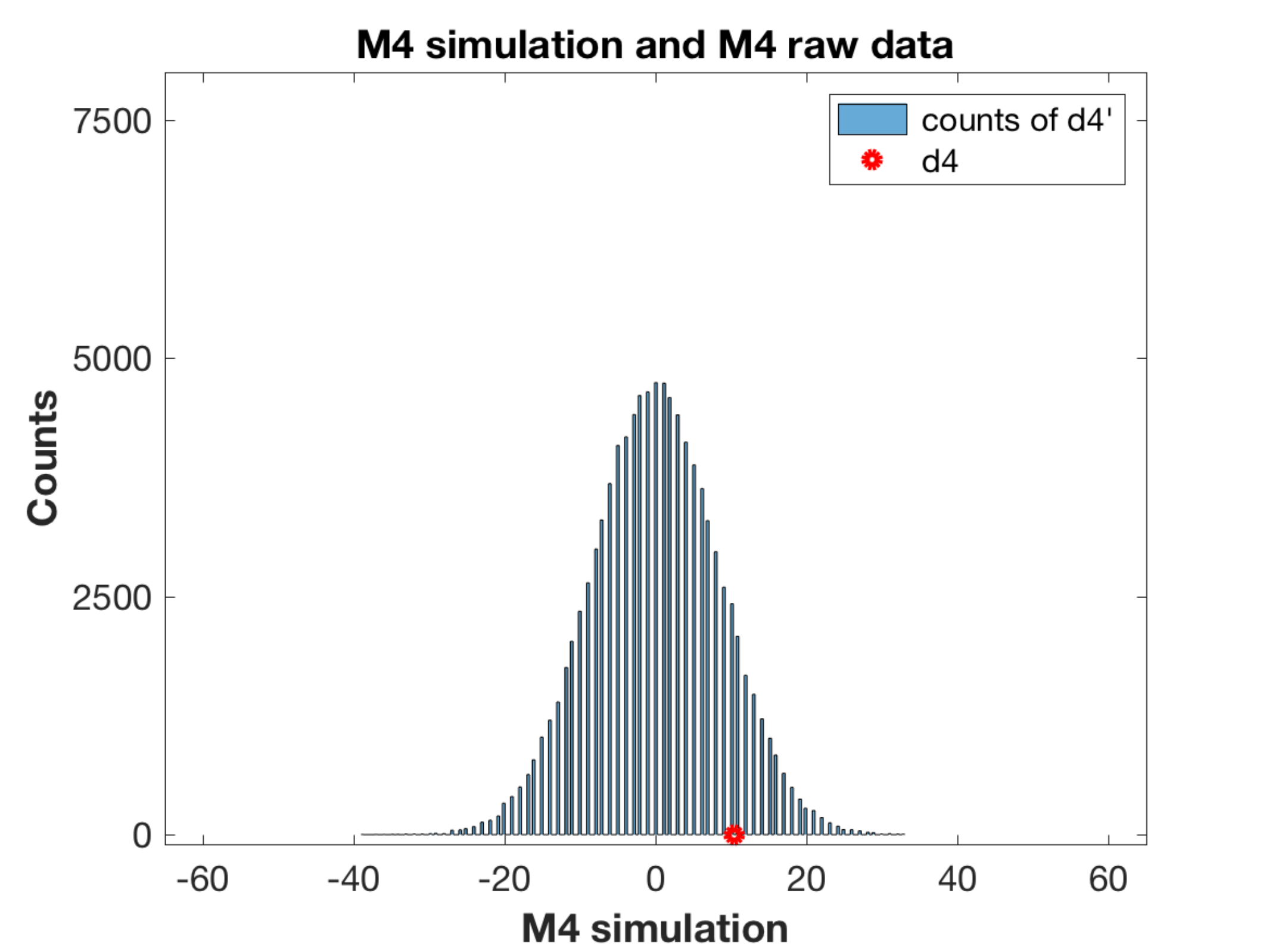}
    \caption{Measure $(M4)$, $\text{p-value} = 0.1078$}
  \end{subfigure}

  \begin{subfigure}[b]{0.5\linewidth}
    \includegraphics[width=.75\linewidth]{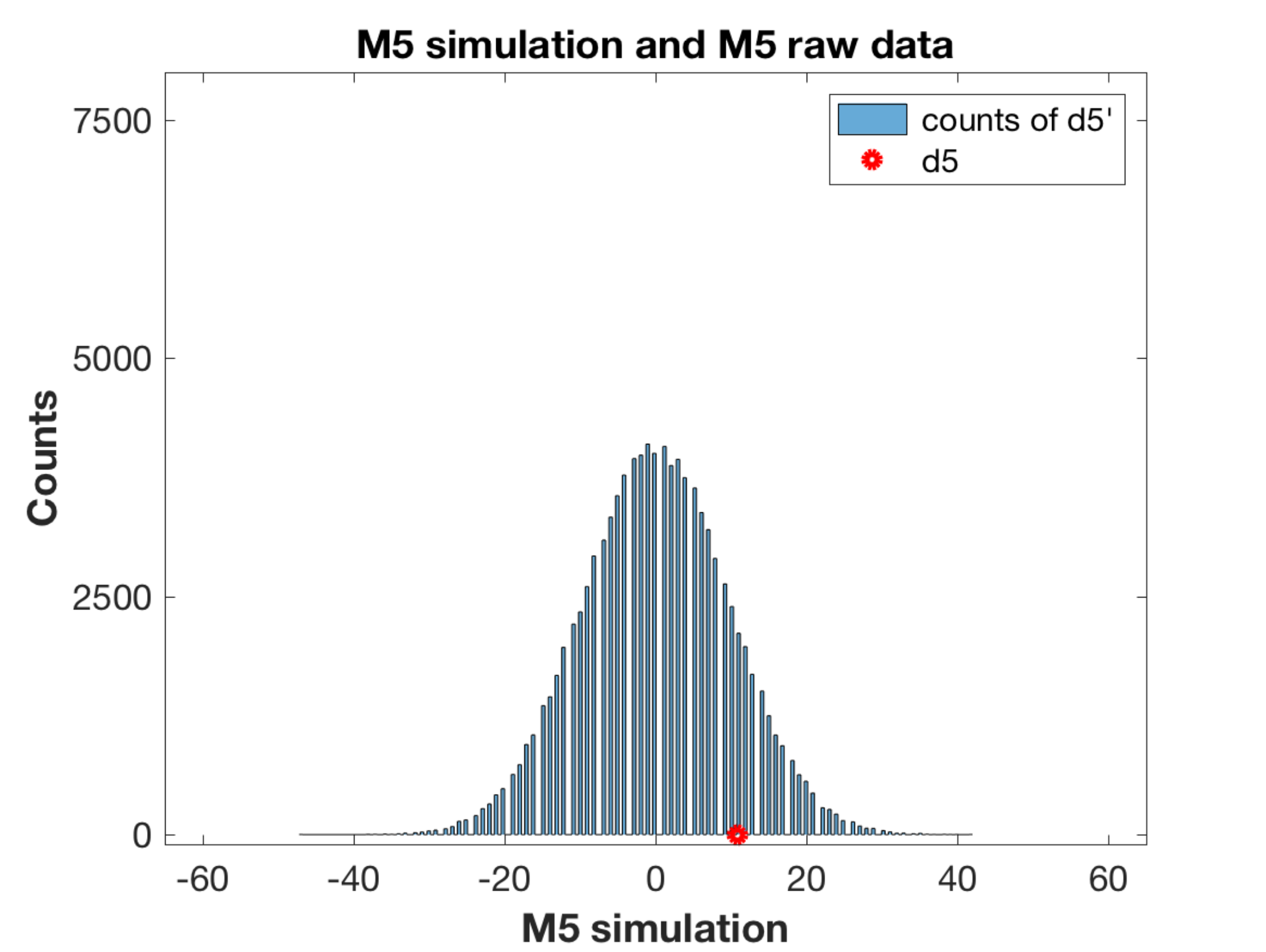}
    \caption{Measure $(M5)$, $\text{p-value} = 0.1428$}
  \end{subfigure}
  \begin{subfigure}[b]{0.5\linewidth}
    \includegraphics[width=.75\linewidth]{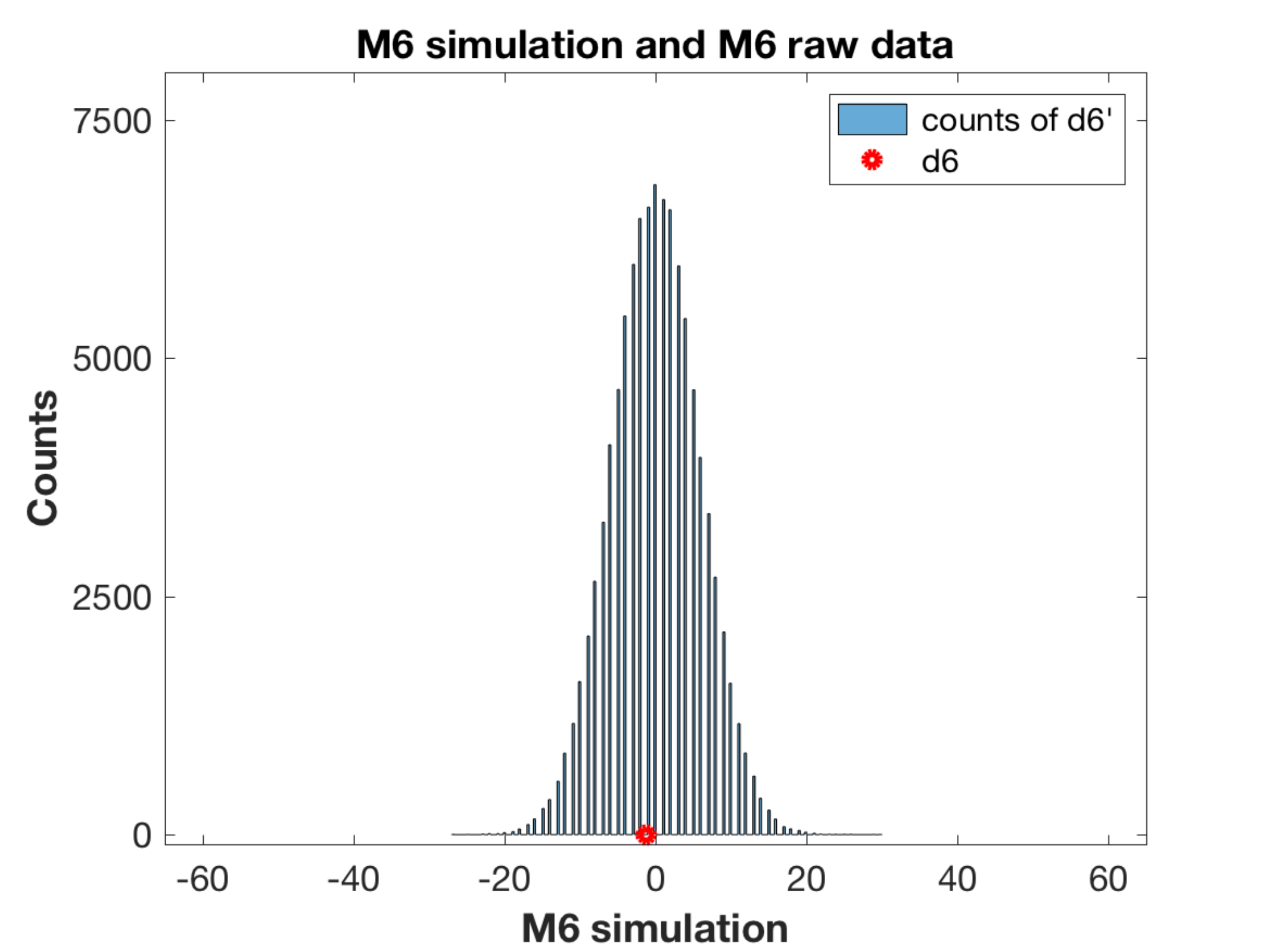}
    \caption{Measure $(M6)$, $\text{p-value} = 0.3987$}
  \end{subfigure}

  \begin{subfigure}[b]{0.5\linewidth}
    \includegraphics[width=.75\linewidth]{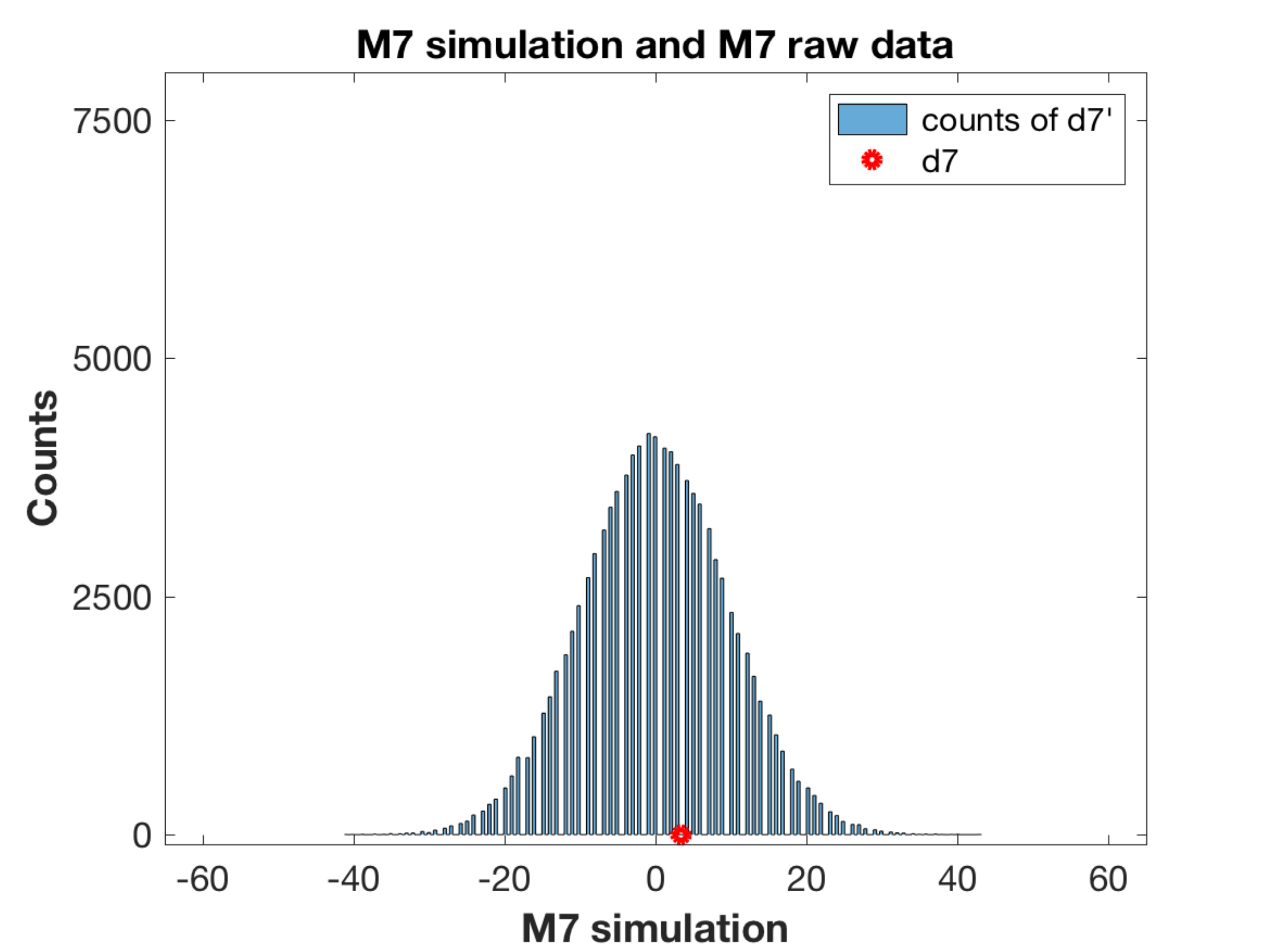}
    \caption{Measure $(M7)$, $\text{p-value} = 0.3583$}
  \end{subfigure}
  \begin{subfigure}[b]{0.5\linewidth}
    \includegraphics[width=.75\linewidth]{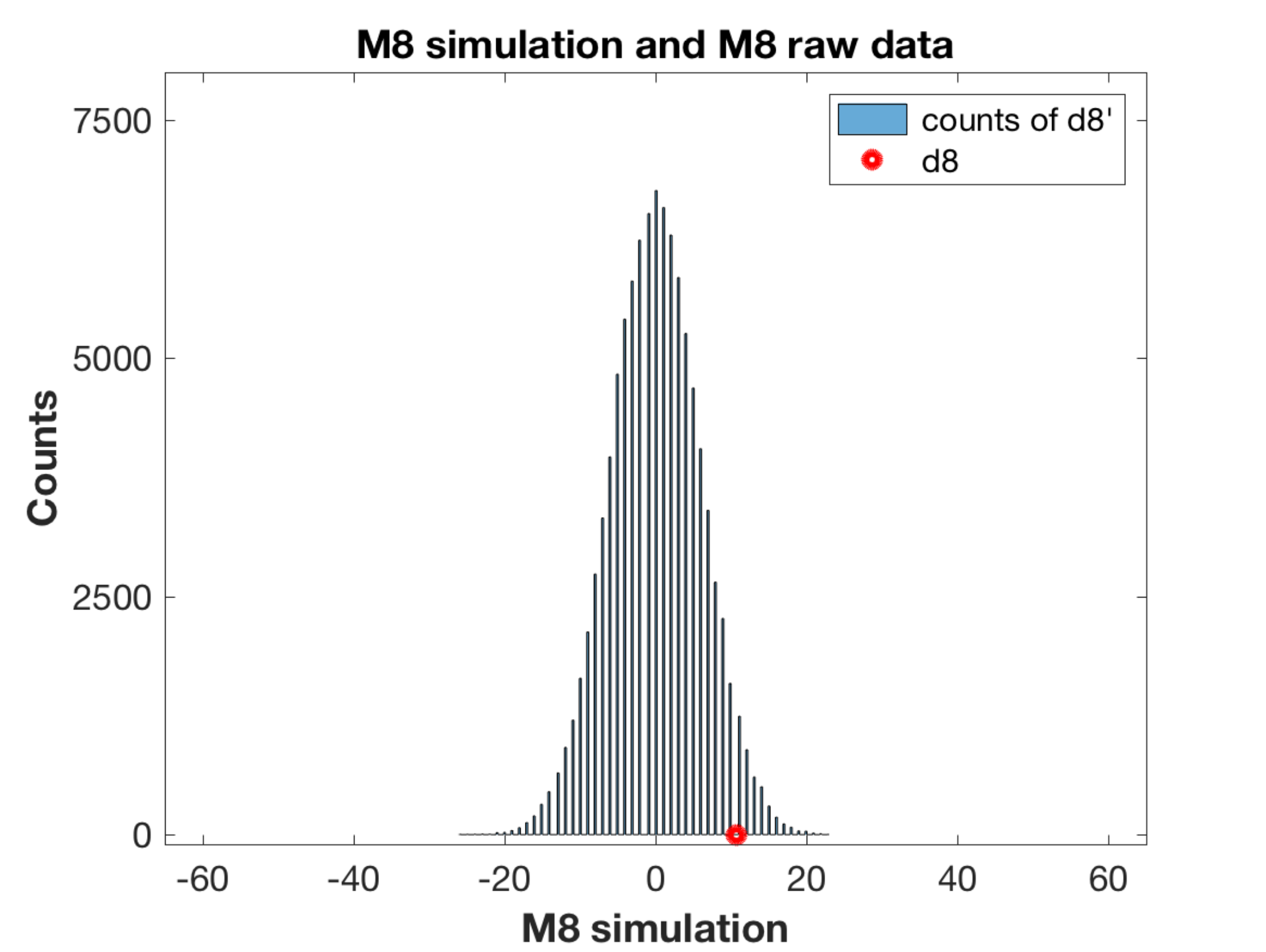}
    \caption{Measure $(M8)$, $\text{p-value} = 0.0396$}
  \end{subfigure}
  \caption{Bootstrapped distribution with 100,000 trials of measures $(M1)-(M8)$ from the simulated data for a plausible representation of a community based intervention.}
\label{robin_data1}
 \end{figure*}

In order to utilize the 64 numbers as raw counts, first, we transform the probabilities into counts by multiplying each probability by 100. Note that the total number of people in the study now becomes 1,600. Applying the bootstrapping technique to answer the eight measures $(M1)-(M8)$ with $B_{N} = 100,000$ times of simulations, we got the p-values of the measures equal to $(0.000,0.003,0.4560,0.1078,0.1428,0.3987,0.3583,0.0396)$ correspondingly (see Fig \ref{robin_data1}).

We also use the analytical method to compute the tail probabilities of the Poisson-Binomial distributions (see Eq \ref{PB-eq}) directly using the MATLAB codes provided by \cite{fernandez2010closed} and get the p-values of the eight measures equal to $(0.0000,0.0003,0.4564,0.1076,0.1425,0.3989,0.3588,0.0398)$ correspondingly, and these are close to the bootstrapped result up to $10^{-3}$.

\end{document}